\input harvmac.tex
%\draftmode
\def\wt{\widetilde}

\def\am{\alpha_{-n}}
\def\ap{\alpha_n}
\def\pmi{\psi_{-r}}
\def\pp{\psi_{r}}
\def\vt{\vartheta}
\def\b{\bar}

\def\vac{|\eta,k\hskip-3pt>}
\def\mvac{|-\eta,k\hskip-3pt>}
\let\includefigures=\iftrue
\newfam\black
\includefigures
\input epsf
\def\figin{\epsfcheck\figin}\def\figins{\epsfcheck\figins}
\def\epsfcheck{\ifx\epsfbox\UnDeFiNeD
\message{(NO epsf.tex, FIGURES WILL BE IGNORED)}
\gdef\figin##1{\vskip2in}\gdef\figins##1{\hskip.5in}% blank space
instead
\else\message{(FIGURES WILL BE INCLUDED)}%
\gdef\figin##1{##1}\gdef\figins##1{##1}\fi}
\def\DefWarn#1{}
\def\figinsert{\goodbreak\midinsert}
\def\ifig#1#2#3{\DefWarn#1\xdef#1{fig.~\the\figno}
\writedef{#1\leftbracket fig.\noexpand~\the\figno}%
\figinsert\figin{\centerline{#3}}\medskip
\centerline{\vbox{\baselineskip12pt
\advance\hsize by -1truein\noindent
\footnotefont{\bf Fig.~\the\figno:} #2}}
\bigskip\endinsert\global\advance\figno by1}
\else
\def\ifig#1#2#3{\xdef#1{fig.~\the\figno}
\writedef{#1\leftbracket fig.\noexpand~\the\figno}%
%\figinsert\figin{\centerline{#3}}\medskip
\centerline{\vbox{\baselineskip12pt
%\advance\hsize by -1truein\noindent
\footnotefont{\bf Fig.~\the\figno:} #2}}
%\bigskip\endinsert
\global\advance\figno by1}
\fi
%
%%%%%%%%% citation macros %%%%%%
%
\def\np#1#2#3{Nucl. Phys. {\bf{B{#1}}} (#2) #3}

\def\cmp#1#2#3{Comm. Math. Phys. {\bf{{#1}}} (#2) #3}

\def\gm#1{{\Gamma\left({#1}\right)}}
\def\dig#1{{\Psi\left({#1}\right)}}
%
%%%%%%%%% macros %%%%%%%%%%%%%%%
% rami's macros
%
%

\def\bc{{\bf C}}

\def\co{{\cal O}}

\def\ka{K\"ahler}

\def\lg{{\hbox{log}}}
%%%%%%%%%%%%%%%%%%%%%%%%%%%%%%%%%%%%%%%%%
\nopagenumbers
\baselineskip 12pt
\leftskip 3.9 in \vbox {hep-th/9906242}
\leftskip 3.9 in \vbox {IASSNS-HEP-99/60}
\leftskip 3.9 in \vbox {RU-99-26}
\leftskip 3.9 in \vbox {SU-ITP-99-31}
\vskip .32 in
\font\bigrm = cmr10 scaled \magstep 3
\centerline{\bigrm Fractional Branes and Boundary States in}
\vskip 15pt
\centerline{\bigrm Orbifold Theories}
\vskip .32in
\leftskip 0 in
\baselineskip 16 pt
\centerline{Duiliu-Emanuel Diaconescu$^\natural$
and Jaume Gomis$^\sharp$}
\vskip .12in
\centerline{\it $^\natural$ School of Natural Sciences, Institute for
Advanced Study}
%\centerline{\it Institute for Advanced Study}
\centerline{\it Olden Lane, Princeton, NJ 08540}
\footnote{${}^{}$}{${}^{\natural}$\tt diacones@sns.ias.edu}
\vskip .05in
\centerline{\it $^\sharp$ Department of Physics and Astronomy, Rutgers
University}
%\centerline{\it Rutgers University}
\centerline{\it Piscataway, NJ 08855-0849}
\centerline{\it and}
\centerline{\it Department of
Physics, Stanford University} \centerline{\it
Stanford, CA 94305-4060}
\footnote{${}^{}$}{${}^{\sharp}$\tt jgomis@leland.stanford.edu}
\smallskip

We study the D-brane spectrum of ${\cal N}=2$ string orbifold
theories using the boundary state formalism.
 The construction
is carried out for orbifolds with isolated singularities, non-isolated
singularities and orbifolds with discrete torsion. Our results
agree with the corresponding K-theoretic predictions when they are
available and generalize them when they are not. This suggests
that the classification of boundary states provides a sort of 
"quantum K-theory"
just as chiral rings in CFT provide "quantum" generalizations of
cohomology. We discuss the identification of these states with D-branes
wrapping holomorphic cycles in the large radius limit of the
CFT moduli space. The
example ${\bf C}^3/{\bf Z}_3$ is worked out in full detail using
local mirror symmetry techniques. We find a precise correspondence
between fractional branes at the orbifold
point and configurations of D-branes
described by vector bundles on the exceptional ${\bf P}^2$ cycle.

\vskip -1in 
\Date\quad
\pageno=1
\baselineskip 16pt

\newsec{Introduction}

The realization that D-branes
\nref\dbranes{J. Dai, R. Leigh, and
J. Polchinski, "New Connections Between String Theories", Mod.
Phys. Lett. {\bf A4} (1989) 2073;\parskip=0pt
\item{}R. Leigh, "Dirac-Born-Infeld Action from Dirichlet Sigma Model",
Mod. Phys. Lett. {\bf A4} (1989) 2767;
\parskip=0pt
\item{}
P. Horava, ``Strings on World-Sheet Orbifolds,'' 
Nucl. Phys. {\bf B327} (1989) 461;
\parskip=0pt
\item{}
P. Horava, "Background Duality of Open String Models", Phys. Lett.
{\bf B231} (1989) 251;\parskip=0pt
\item{}
M.B. Green, "Space-time Duality and Dirichlet String Theory", Phys.
Lett. {\bf B266} 325 (1991),
"Pointlike States for  Type IIB Superstrings",
Phys. Lett. {\bf B329} (1994) 435, hep-th/9403040;
 "A Gas of D Instantons", Phys.
Lett. {\bf B354} (1995) 271, hep-th/9504108;\parskip=0pt
\item{}
J. Polchinski, "Combinatorics of Boundaries in String Theory",
Phys. Rev. {\bf D50} (1994) 6041.}%
\dbranes\ carry Ramond-Ramond
charge \nref\polch{J. Polchinski, "Dirichlet-Branes and
Ramond-Ramond Charge", Phys. Rev. Lett. {\bf 75} (1995) 4724,
hep-th/9510017.}%
\polch\
has been a crucial ingredient in recent
progress in nonperturbative string theory. The simple CFT
description of these states as hyperplanes on which open strings
can end has provided new insights in the structure of space-time
at substringy length scales
\nref\dkps{M.R. Douglas, D. Kabat, P.
Pouliot and  S.H. Shenker, "D-branes and Short Distances in String
Theory", Nucl. Phys. {\bf B485} (1997) 85, hep-th/9608024.}%
\dkps\
and has given  an adequate framework  for attempting a
microscopic formulation
\nref\bfss{T. Banks, W. Fischler, S.H.
Shenker and  L. Susskind, "M Theory As A Matrix Model: A
Conjecture", Phys. Rev. {\bf D55} (1997) 5112, hep-th/9610043.}%
\bfss\
of M-theory. However, there are
many vacua of string theory which do not admit a clear space-time
interpretation --they are described by an abstract CFT -- and for
which  the description of D-branes is not manifest. Moreover,
these vacua are expected to contain nonperturbative states
carrying Ramond-Ramond charge and to fit in the web of dualities
relating the different string theories.  It is therefore important
to describe D-branes in general string vacua and at generic points
of moduli space, where a space-time interpretation might not be
obvious, in order to unravel the complete description of the physics
of M-theory.

A powerful tool for analyzing the inclusion of boundaries on
string worldsheets is the boundary state formalism
\nref\bsta{J.
Polchinski and Y. Cai, "Consistency of Open Superstring Theories",
Nucl. Phys. {\bf B296} (1988) 91;\parskip=0pt
\item{}
C. Callan, C. Lovelace, C. Nappi and S. Yost, "Loop Corrections to
Superstring Equations of  Motion", Nucl. Phys. {\bf B308}  (1988)
221;\parskip=0pt
\item{}
T. Onogi and N. Ishibashi,
 ``Conformal Field Theories On Surfaces With Boundaries And
 Crosscaps",
Mod. Phys. Lett. {\bf A4} (1989) 161;
\parskip=0pt
\item{}
N. Ishibashi,
 ``The Boundary And Crosscap States In Conformal Field Theories",
Mod. Phys. Lett. {\bf A4} (1989) 251.}%
\bsta .
This formalism
provides  a closed string description of D-branes and it is
applicable, in principle, to arbitrary CFTs. In this vein, one can
tackle the question of determining the D-brane spectrum at an
arbitrary point of moduli space of string theory. Along different
lines, Witten
\nref\W{E. Witten, ``D-Branes and K Theory'', JHEP
{\bf 12} (1998) 019, hep-th/9810188.}%
\W\ has argued that the
classification of D-branes in a general space-time background has
to be upgraded from singular homology to K-theory.
This generalization has brought to light a new understanding of
D-branes and the appearance of new  hitherto unsuspected states.
We strongly believe that the classification of boundary states of
a given CFT provides a stringy generalization of K-theory and that
its study should provide a sort of "quantum  K-theory" just as
CFT chiral rings generalized cohomology to "quantum cohomology"
\nref\orbcft{L. Dixon, J.A. Harvey, C. Vafa and  E. Witten,
"Strings on Orbifolds", Nucl. Phys. {\bf B261} (1985) 678, "Strings
on Orbifolds 2", Nucl. Phys. {\bf B274} (1986) 285.}%
\orbcft . Some hints of this will appear in this work.
Moreover, the understanding of D-branes at a generic point in
moduli space and D-geometry should shed new light on what
space-time at the shortest length scales really is.

In this paper we classify D-brane states at certain points of the
moduli space of Type II strings with eight supercharges. In
particular we provide a boundary state description of D-brane
states at points in moduli space that admit a perturbative
orbifold CFT description\foot{Some previous work on boundary
states and orbifolds can be found in \nref\orb{ H. Ooguri, Y. Oz
and  Z. Yin, "D-Branes on Calabi-Yau Spaces and Their Mirrors",
Nucl. Phys. {\bf B477} (1996) 407, hep-th/9606112; \parskip=0pt
\item{} F. Hussain, R. Iengo, C. Nunez and C. Scrucca,
"Interaction of moving D-branes on orbifolds", Phys. Lett. {\bf
B409} (1997) 101, hep-th/9706186;\parskip=0pt \item{}
J. Fuchs, C. Schweigert, "Branes: from free fields to general backgrounds", Nucl. 
Phys. {\bf B530} (1998) 99,
hep-th/9712257,
\parskip=0pt \item{}
M.
Bertolini, R. Iengo and C. Scrucca, "Electric and magnetic
interaction of dyonic D-branes and odd spin structure", Nucl. Phys.
{\bf B522} (1998) 193, hep-th/9801110; \parskip=0pt \item{} M.
Bertolini, P. Fre, R. Iengo and C. Scrucca, "Black holes as
D3-branes on Calabi-Yau threefolds", Phys. Lett. {\bf B431} (1998)
22, hep-th/9803096;
\parskip=0pt \item{}
O. Bergman and M.R. Gaberdiel, "Stable non-BPS D-particles",
Phys. Lett. {\bf B441} (1998) 133, hep-th/9806155;\parskip=0pt
\item{} O. Bergman and M.R. Gaberdiel, "Non-BPS States in
Heterotic - Type IIA Duality", JHEP {\bf 03} (1999) 013;
hep-th/9901014;\parskip=0pt \item{} M.Bill\'o, B. Craps and F.
Roose, "On D-branes in Type 0 String Theory",
hep-th/9902196;\parskip=0pt \item{} I. Brunner, A. Rajaraman and
M. Rozali, "D-branes on Asymmetric Orbifolds";
hep-th/9905024;\parskip=0pt \item{} I. Brunner, R. Entin and C.
R\"omelsberger, "D-branes on $T^4/Z_2$ and T-Duality", JHEP {\bf
06} (1999) 016, hep-th/9905078.}%
\orb .}. This gives a closed
string description of the  fractional D-branes in
\nref\Da{M.R.
Douglas, ``Enhanced Gauge Symmetry in M(atrix) Theory'', JHEP {\bf
07} (1997) 004, hep-th/9612126.}%
\nref\DGM{M.R. Douglas, B.R
Greene and D.R. Morrison, "Orbifold Resolution by D-Branes",
Nucl. Phys. {\bf B506} (1997) 84, hep-th/9704151.}%
\nref\DDG{D.-E.
Diaconescu, M.R. Douglas and J. Gomis, "Fractional Branes and
Wrapped Branes", JHEP {\bf 02} (1998) 013.}%
\nref\diagom{D.-E.
Diaconescu and J. Gomis, "Duality in Matrix Theory and Three
 Dimensional Mirror Symmetry", Nucl. Phys. {\bf B517} (1998) 53,
 hep-th/9707019.}%
\refs{\Da,\DGM,\DDG,\diagom}. As a byproduct  of the construction
of consistent boundary states, we give a physical  D-brane
proof of the McKay correspondence
\nref\McK{J. McKay, "Graphs, Singularities and Finite Groups",
Proc. Symp. in Pure Math. {\bf 37} (1980) 183.}%
\nref\GSV{C.
Gonzalez-Springer and  J. Verdier, ``Construction G\'eom\'etrique de
la Correspondance de McKay'', Ann. Scient. Ec. Norm. Sup. (1983),
409.}%
\nref\AV{M. Artin and J. Verdier, ``Reflexive modules over
rational double points, Math. Ann. {\bf 270} (1985), 79.}%
\nref\BD{V.V. Batyrev and D.I. Dais, ``Strong McKay
Correspondence, String-theoretic Hodge Numbers and Mirror
Symmetry'', Topology {\bf 35} (1996) 901, alg-geom/9410001.}%
\nref\Bryl{J.-L. Brylinski, ``A Correspondence Dual to McKay's'',
alg-geom/9612003.}%
\nref\IR{Y. Ito and M. Reid, ``The
McKay Correspondence for Finite Subgroups of $SL(3,{\bf C})$'',
Higher Dimensional Complex Varieties
(Trento 1994), 221-240, de Gruyter, Berlin, 1996, alg-geom/9411010.}%
\nref\R{M. Reid. ``McKay Correspondence'',
alg-geom/9702016.}%
\nref\IN{Y. Ito and H. Nakajima, ``McKay
Correspondence and Hilbert Schemes in Dimension Three'',Proc. Japan.
Acad. Ser. A. Math. Sci. {\bf 72} (1996) 172, alg-geom/9803120.}%
\refs{\McK-\IN}. Roughly speaking, this correspondence states that
there is a one-to-one correspondence between the number of
non-trivial irreducible representations of the orbifold group
$\Gamma$ describing the isolated orbifold singularity ${\bf
C}^d/\Gamma$ and the homology generators of the crepant resolution
of the singularity. In section 5 this will be explained in the
more precise language of K-theory. In our physical situation, this
follows from a deformation of the BPS spectrum determined by
boundary states to the large radius limit of the orbifold moduli
space where the states can be realized as D-branes wrapping
supersymmetric cycles. We also analyze boundary states for certain
non-isolated singularities and find that D-brane considerations
suggest a generalization of the McKay correspondence to this case.
This would relate non-trivial irreducible representations of the
discrete group to compact homology generators of the resolved
space.

The reduced amount of supersymmetry of these vacua introduces
corrections to the special geometry of the moduli space,
which makes the abovementioned identification very difficult.
In section $5$ we   work out an example by providing
the exact solution for
the ${\bf C}^3/{\bf Z}_3$ model using the techniques of local mirror
symmetry \nref\A{P.S. Aspinwall, ``Resolution of Orbifold
Singularities in String Theory'', Mirror Symmetry II, B.R. Greene
and S.T. Yau eds, hep-th/9403123.}%
\nref\KKV{S. Katz, A. Klemm
and C. Vafa, ``Geometric Engineering of Quantum Field Theories'',
\np{497}{1997}{173}, hep-th/9609239.}%
\nref\KMV{S. Katz, P. Mayr
and C. Vafa, ``Mirror Symmetry and Exact Solution of $4D$ $N=2$
Gauge Theories -- I'', Adv. Theor. Math. Phys. {\bf 1} (1998) 53,
hep-th/9706110.}%
\nref\CKYZ{T.-M. Chiang, A. Klemm, S.-T. Yau and
E. Zaslow, ``Local Mirror Symmetry: Calculations and
Interpretations'', hep-th/9903053.}%
\refs{\A, \KKV,\KMV,\CKYZ}.
We identify the singularities in moduli space, the monodromies
around these points and the exact central charge for the model.
This allows us to make a precise identification between boundary
states at the orbifold point with branes in the large volume
limit. Moreover, the  fractional branes at the
orbifold are extended to an arbitrary point in moduli space.
Similar results for the quintic moduli space have been obtained in
\ref\BD{I. Brunner, M.R. Douglas, A. Lawrence and C. R\"omelsberger,
``D-Branes on The Quintic'', hep-th/9906200.}. The role of the
perturbative orbifold point is played there by the Gepner point.

The boundary state formalism serves very useful in analyzing the
spectrum of models that do not have an obvious geometrical
interpretation such as orbifolds with discrete torsion \nref\V{C.
Vafa, "Modular Invariance and  Discrete Torsion on Orbifolds",
Nucl. Phys. {\bf B273} (1986) 592.}%
\nref\VW{C. Vafa and E. Witten,
"On Orbifolds with Discrete Torsion", J. Geom. Phys. {\bf 15} (1995)
189, hep-th/9409188.}%
\refs{\V,\VW}. We construct the consistent
set of boundary states corresponding to fractional branes for
models with discrete torsion and find that these are classified by
irreducible projective representations of the discrete group.
Since orbifolds in this class do not admit complete smooth
resolutions, there is no obvious geometric interpretation similar
to McKay correspondence.
The ease by which the boundary states generalize to these models
indeed suggest that the most general framework for D-branes is the
boundary state formalism. In certain situations, this conformal
field theory construction reduces to a known mathematical structure
such as cohomology or K-theory. We believe that this is also the
case for orbifolds with discrete torsion, where the fractional branes
should be described by an appropriate generalization of equivariant
K-theory \refs{\W}.

The paper is organized as follows. In section $2$ we summarize the
description of fractional branes at orbifold singularities and
some of their properties from a D-brane probe perspective. In section
$3$ we give a selfcontained discussion of boundary states and
construct the boundary states in flat space and set up the
notation. In section $4$ we give the general prescription required
to write down boundary states for perturbative  orbifold CFTs and
give a general argument which predicts the number of states. We
then go on and explicitly work out examples involving orbifolds
with isolated singularities, with non-isolated singularities and
with discrete torsion. In section $5$ we elucidate the role of
these boundary states and their relation to D-brane states at the
large volume limit, and explain how the boundary state construction
sheds new
light onto the McKay correspondence. In section $6$ we consider the
${\bf C}^3/{\bf Z}_3$ example and work out the exact solution of the
model including worldsheet instanton corrections. We then make a
precise mapping between fractional branes  and D-branes in the large
volume limit. We also extend these states to generic points in moduli
space. In the Appendix we
summarize conventions and some useful formulas.

\newsec{Branes at Orbifold Singularities and Fractional Branes}

The gauge theory on a D-brane probe reproduces the space-time in
which it is embedded as its moduli space of vacua. It also
captures many features of the space-time BPS states which
correspond to wave functions on the moduli space of the brane.
When the brane theory is known, this can be an efficient tool in
addressing dynamical problems. Hence, in these situations, it is
very useful to establish an explicit correspondence between field
theory and space-time excitations. A very rich background for
testing these ideas is given by orbifold singularities with a
perturbative conformal field theory description. In this section
we will give a brief presentation of D-particle states in orbifold
theories of the form ${\bf C}^d/\Gamma$, $\Gamma\subset SU(d)$,
from the probe perspective 
\nref\DM{M.R. Douglas and G. Moore, "D-branes, Quivers, and
ALE Instantons", hep-th/9603167.}%
\nref\JM{C.V. Johnson and R.C. Myers, 
``Aspects of Type IIB Theory on ALE Spaces'', Phys. Rev. 
{\bf D55} (1997) 6382, hep-th/9610140.}%
\refs{\DM,\JM}. The following sections will contain a
more detailed description of the same states using the boundary
state formalism.

The theory on D-branes probing a ${\bf C}^d/\Gamma$ singularity is
uniquely determined by choosing a representation of \foot{We will
consider supersymmetric backgrounds so that $\Gamma$ is a discrete
subgroup of $SU(d)$.} $\Gamma$, which defines the action on the
Chan-Paton indices, and by specifying an action of $\Gamma$ on
${\bf C}^d$ via a $d$-dimensional representation $R$.
 The bosonic projection equations that need to be solved are
\eqn\projection{\eqalign{ \gamma(g_i)A\gamma(g_i)^{-1}&=A\qquad
\qquad\qquad i=1,\ldots,|\Gamma|\cr
\gamma(g_i)Z^{\alpha}\gamma(g_i)^{-1}&=R(g_i)^{\alpha}_{\beta}Z^{\beta}\qquad
\alpha=1,\ldots,d},} where $Z^\alpha$ are the complex fields which
parameterize ${\bf C}^d$.
 $\gamma$ is any representation of the orbifold
group $\Gamma$ \eqn\representation{
\gamma=\oplus^{n}_{a=1}N_a\gamma_a,} consisting of $N_a$ copies of
the $a$-th irreducible representation $\gamma_a$ of $\Gamma$. The
gauge group of the D-brane theory is the commutant of $\gamma$ in
$U(N)$, where $N=\sum_{a=1}^{n}d_aN_a$ with
$d_a=\hbox{dim}(\gamma_a)$, which yields a
$G=\prod_{a=1}^{n}U(N_a)$ gauge theory. The matter content can be
similarly found and it is encoded in the representation theory of
$\Gamma$. The gauge theory describes the physics of branes at the
singularity.

Although this formalism is very general, in the following we
specialize to D0-brane quantum mechanics. In the Born Oppenheimer
approximation, the moduli space of vacua has two branches with
different space-time interpretation. The Higgs branch -- spanned
by the expectation values of the fields $Z^\alpha$ -- reproduces
the orbifold geometry. From a dynamical point of view, excitations
propagating along this branch correspond to D0-branes that can
move about the singularity. The theory also has a Coulomb branch
parameterized by expectation values of the fields $\phi^i$ in the
vector multiplet, the $Z^\alpha$ being set to zero. The
excitations along the Coulomb branch have been interpreted in
\refs{\Da,\DGM,\DDG,\diagom} as branes wrapping the shrunken cycles of the
orbifold singularity. There is a one-to-one correspondence between such
BPS states and irreducible representations of the orbifold group
$\Gamma$. If we resolve the orbifold singularity by turning on
marginal operators, the same states can be identified with branes
wrapping supersymmetric cycles in a smooth space-time background.
This can be thought of as a physical McKay correspondence and it
will be considered in more detail later.

These states carry charge under the untwisted, as well as twisted
RR fields. The values of the charges can be determined by an open
string disk computation as outlined in
\refs{\DM,\DDG}. The state corresponding to the $a$-th irreducible
representation has charge \eqn\fract{ Q^a_0={d_a\over|\Gamma|}\qquad a=1,\ldots,n}
with respect to the untwisted RR field and \eqn\twistch{
Q^a_m={\chi^a(g_m)\over |\Gamma|}\qquad a=1,\ldots,n} with respect to the $m$-th
twisted RR field, where $\chi^a(g_m)=\hbox{Tr}(\gamma_a(g_m))$ is
the character of the $a$-th representation. Note that the
D0-brane charge \fract\ is fractional.

A similar analysis can be made for orbifolds with discrete
torsion. As shown in
\nref\Db{M.R. Douglas, "D-branes and Discrete Torsion", hep-th/9807235.}%
\nref\DF{M.R. Douglas and B. Fiol, "D-Branes and Discrete Torsion
II", hep-th/9903031.}%
\refs{\Db,\DF}, discrete torsion can  be incorporated in the probe
gauge theory by considering Chan-Paton factors in a  projective
representation of the gauge group. The cocycle factors appearing
in the orbifold conformal field theory partition function
\refs{\V,\VW} are representatives of a cohomology class in
$H^2(\Gamma,U(1))$.

It can be shown using the methods of \ref\Karp{G. Karpilowsky,
"The Schur Multiplier", London Mathematical Society Monographs.
New Series, 2. The Clarendon Press, Oxford University Press,
New York, 1987.}, that all
discrete subgroups of $SU(2)$ have trivial $U(1)$-valued
cohomology, so discrete torsion cannot be implemented in ${\bf
C}^2/\Gamma$ orbifolds\foot{This can also be easily seen by
showing that all discrete subgroups of $SU(2)$ do not admit
non-trivial projective representations and taking into account the
relation between $H^2(\Gamma,U(1))$ and classes of projective
representations mentioned below.}. Nontrivial examples can be
found for $\Gamma\subset SU(3)$ acting on ${\bf C}^3$. Given
$\Gamma\in SU(3)$ with non-trivial $H^2(\Gamma,U(1))$, and  a set
of cocycle factors defining the discrete torsion, the gauge theory
on the D-brane probe is found by solving \projection\ with a
projective action of the orbifold group on Chan-Paton factors.
The choice of a cohomology class in $H^2(\Gamma,U(1))$ --
specifying the discrete torsion phases -- determines a class
of irreducible projective representations of $\Gamma$ modulo
projective equivalence.
Within this class, there
are in general $m\neq 1$ (linear equivalence classes of)
irreducible projective representations.
These determine the gauge group to be $G=\prod_{b=1}^m U(N_b)$ just
as before. An important difference here is that the number of
projective representations, $m$, is not linked anymore with the
number of conjugacy classes of $\Gamma$. The gauge theory probe
has again $m$ BPS states on the Coulomb branch. However, since the
resolution of these orbifolds is not complete in string theory
\refs{\VW,\Db,\DF}, these states cannot be given a clear geometric
interpretation. We believe that they can be given nevertheless an
appropriate boundary state and K-theoretic interpretation, but we
will return to this later on.

In the next sections we will construct all these states using the
boundary state formalism  and reproduce their known properties
extracted from the probe theory approach.

\newsec{D-branes as Boundary States}

In the next  section we will construct boundary states
corresponding to the fractional branes introduced in the previous
discussion. For completeness we  introduce  here the construction of
boundary states in flat space and set up the notation.

\subsec{General Construction}

 Boundary states describe in  the closed string theory language the
inclusion
of boundaries on the worldsheet \bsta . On these boundaries, conformally
invariant boundary conditions are imposed on the worldsheet closed
string fields. The eigenstates of these boundary conditions are
called Ishibashi states. Consistency of factorization with the
open string channel
\nref\cardy{J.L. Cardy, "Boundary conditions, fusion rules and the
Verlinde formula, Nucl.Phys {\bf B324} (1989) 581.}%
\cardy\
select  appropriate linear combinations of
Ishibashi states, which are called boundary states.
Boundary
states provide a complementary description of D-branes, which
have a  simple open string description when a space-time
interpretation is available. It is therefore important to analyze
boundary states for generic CFTs since many vacua of string theory
do not admit a clear space-time interpretation.

The non-linear constraints imposed by conformal invariance on
Ishibashi states are usually solved by replacing them by linear
constraints. The
cylinder amplitude describing propagation of a boundary state into
another must admit an open string interpretation as a one loop
vacuum amplitude with integer degeneracy of states at any given
mass level. This constraint, usually referred as Cardy's
condition \cardy , severely restricts the allowed linear combinations of
Ishibashi states.

For a general CFT with a boundary, conservation of momentum across
the boundary imposes the following condition
\eqn\momcons{ T_L(t,\sigma)\vert_{t=0}=T_R(t,\sigma)\vert_{t=0},}
 which
relates the left-moving and right-moving stress energy tensors of
the "bulk" CFT at the $t=0$ boundary. For a CFT with a more
general left-right symmetric chiral algebra $\cal{A}$ than the Virasoro
algebra, one may identify the extra symmetry generators on the left
with those on
the right at the boundary up to the action of an automorphism of
$\cal{A}$\foot{For a nice discussion on this see
\nref\reckscho{A. Recknagel, V. Schomerus, "D-branes in Gepner models", Nucl. 
Phys. {\bf B531} (1998)
185, hep-th/9712186.}%
\reckscho .}.
For the CFT describing string theory in flat space-time, vanishing
of momentum along the boundary requires\foot{$\partial_\pm$ refers
to derivatives with respect to world-sheet light-cone coordinates
$t^\pm=t\pm\sigma$.}
\eqn\mom{
T_{t\sigma}\vert_{t=0}=\partial_tX^{M}\partial_{\sigma}X_{M}
+i\wt\psi^M\partial_+\wt\psi_M-i\psi^M\partial_-\psi_M\vert_{t=0}=0,
\qquad M=0,\ldots,9}
which is satisfied by imposing the linear conditions
\eqn\bc{\eqalign{
\partial_tX\vert_{t=0}&=0 \Longrightarrow (\partial_-X+\partial_+X)
\vert_{t=0}=0\qquad \hbox{Neumann}\cr
\partial_\sigma X\vert_{t=0}&=0 \Longrightarrow
(\partial_- X-\partial_+X)\vert_{t=0}=0
\qquad \hbox{Dirichlet}},}
corresponding to Neumann and Dirichlet boundary conditions
respectively. Superconformal invariance on the worldsheet also relates
the left moving supersymmetry generator with the right moving one
\eqn\super{ (T^F_L-i\eta T^F_R)\vert_{t=0}=0,}where $\eta=\pm 1$
labels the
spin structure. Since $T_L^F=\psi^M\partial_-X_M$,
$T_R=\wt\psi^M\partial_+X_M$ and \bc\ the Neumann and Dirichlet
boundary conditions on worldsheet fermions are
\eqn\world{\eqalign{
(\psi+i\eta\wt\psi)\vert_{t=0}&=0\qquad\hbox{Neumann}\cr
(\psi-i\eta\wt\psi)\vert_{t=0}&=0\qquad\hbox{Dirichlet}},}
which also solve \mom .

We will now write down the boundary states corresponding to D-branes
in the Type II theories and introduce the necessary ingredients to
build boundary states
for orbifolds\foot{We will a notation very similar to that in
\nref\sen{A. Sen, "Stable Non-BPS Bound States of BPS D-branes", JHEP {\bf 9808}
 (1998) 010, hep-th/9805019.}%
\nref\bergabi{O. Bergman and M.R. Gaberdiel, "A Non-Supersymmetric Open String 
Theory and
S-Duality", Nucl.Phys. {\bf B499} (1997) 183, hep-th/9701137.}%
\refs{\sen,\bergabi}.}. For simplicity, we will fix the light-cone gauge by
taking as $x^8\pm x^9$ as light-cone coordinates
\nref\lightcone{M.B. Green and  M.Gutperle, "Light-cone
supersymmetry and D-branes", Nucl.Phys. {\bf B476} (1996) 484,
hep-th/9604091.}%
\lightcone\ after a double
Wick rotation on $x^0$ and $x^8$. The boundary conditions on the
closed string world-sheet fields for a Dp-brane are\foot{See
appendix for conventions.}
\eqn\bound{\eqalign{
\partial_t X^\mu(t=0,\sigma)\vac_{NSNS \atop RR}&=0
\Longrightarrow(\alpha^\mu_{n}+\wt\am^\mu)\vac_{NSNS \atop RR}=0\quad
\mu=0,\ldots,p\cr
\partial_{\sigma}X^i(t=0,\sigma)\vac_{NSNS \atop RR}&=0
\Longrightarrow(\alpha^i_{n}-\wt\am^i)\vac_{NSNS \atop RR}=0 \quad
i=p+1,\ldots,7\cr (\psi^\mu+i\eta\wt\psi^\mu)\vac_{NSNS \atop
RR}&=0 \Longrightarrow(\psi^\mu_{r}+i\eta\wt\pmi^\mu)\vac_{NSNS
\atop RR}=0\cr (\psi^i-i\eta\wt{\psi}^i)\vac_{NSNS \atop RR}&=0
\Longrightarrow(\psi^i_{r}-i\eta\wt\pmi^i)\vac_{NSNS \atop
RR}=0\cr x^{1,2}\vac_{NSNS \atop RR}&=0},} where $\vac_{NSNS \atop
RR}$ carries momentum in the $9-p$ Dirichlet directions\foot{Strictly
speaking the Dp-brane is obtained only after analytically
continuing back.}. These
equations can be easily solved via coherent states
\eqn\coherent{\eqalign{ \vac_{NSNS\atop
RR}=\exp&\Biggl(\sum_{n=1}^{\infty}{1\over
n}(-\sum_{\mu=0,\ldots,p}\am^\mu\wt\am^\mu
+\sum_{i=p+1}^{7}\am^i\wt\am^i)\cr
&+i\eta\sum_{r>0}(-\sum_{\mu=0,\ldots,p}\pmi^\mu\wt\pmi^\mu
+\sum_{i=p+1}^{7}\pmi^i\wt\pmi^i) \Biggl)\vac^{(0)}_{NSNS\atop
RR}}} using the left moving and right moving oscillators of the
string. The Fock vacuum $\vac^{(0)}_{NSNS\atop RR}$ is unique in
the NSNS sector and it is identical to the usual closed string
vacuum. The RR vacuum is more complicated and we will now describe
it
in a way which will be  useful in the orbifold construction.
These vacua must also solve \bound\ for the zero modes. It
turns out to be convenient to rewrite the zero mode equations in
the creation-annihilation basis of the $SO(8)$ Clifford algebra
$\Gamma^{a\pm}={1\over 2}(\gamma^{2a}\pm i \gamma^{2a+1})$, which
satisfy \eqn\cliff{ \{\Gamma^{a+},\Gamma^{b-}\}=\delta^{ab} \qquad
\left(\Gamma^{a+}\right)^2=\left(\Gamma^{a-}\right)^2=0 \qquad
a=0,\ldots,3.} For $p$ odd the equations become
\eqn\zeromode{\eqalign{
(\Gamma^{b-}+i\eta\wt\Gamma^{b-})\vac_{RR}^{(0)}&=0\qquad
b=0,\ldots,{p-1\over 2}\cr
(\Gamma^{c-}-i\eta\wt\Gamma^{c-})\vac_{RR}^{(0)}&=0\qquad
c={p+1\over 2},\ldots,3},} and for $p$ even
\eqn\zeromodeb{\eqalign{
(\Gamma^{d-}+i\eta\wt\Gamma^{d-})\vac_{RR}^{(0)}&=0\qquad
d=0,\ldots,{p\over 2}-1\cr (\Gamma^{{p\over
2}-}+i\eta\wt\Gamma^{{p\over 2}+})\vac_{RR}^{(0)}&=0\cr
(\Gamma^{e-}-i\eta\wt\Gamma^{e-})\vac_{RR}^{(0)}&=0\qquad
d={p\over 2}+1,\ldots,3},} Hence the solution to these equations
are \eqn\boundb{\eqalign{ \vac_{RR}^{(0)}&=\exp\left(i\eta(
-\Gamma^{b+}\wt\Gamma^{b-}+\Gamma^{c+}\wt\Gamma^{c-})\right)
|0,k\hskip-3pt>_{RR}\otimes
\wt{|0,k\hskip-3pt>}_{RR}\qquad \qquad \quad \ \ \hbox{p odd}\cr
\vac_{RR}^{(0)}&=\exp\left(i\eta(
-\Gamma^{d+}\wt\Gamma^{d-}-\Gamma^{{p\over 2}+}\wt\Gamma^{{p\over
2}+} +\Gamma^{e+}\wt\Gamma^{e-})\right)|0,k\hskip-3pt>_{RR}\otimes
\wt{|0,k\hskip-3pt>}_{RR}\ \ \hbox{p even}.}} The Fock vacua are
defined such that
\eqn\fock{ \Gamma^{a-}|0,k\hskip-3pt>_{RR}=0\qquad a=0,\ldots,3}
and \eqn\fockb{\eqalign{
\wt\Gamma^{a+}\wt{|0,k\hskip-3pt>}_{RR}&=0\qquad a=0,\ldots,3\
\qquad \qquad \qquad \qquad \qquad \qquad \hbox{For $p$ odd.}\cr
\wt\Gamma^{d+}\wt{|0,k\hskip-3pt>}_{RR}&=0\qquad
\wt\Gamma^{{p\over 2}-}\wt{|0,k\hskip-3pt>}_{RR}=0\qquad
\wt\Gamma^{e+}\wt{|0,k\hskip-3pt>}_{RR}=0\quad\hbox{For $p$
even.}},} where the indices are as in \zeromode . These are
easily described as vectors in the $SO(8)$ weight lattice in the
standard way.
They
are
\eqn\vect{ |0,k\hskip-3pt>_{RR}\rightarrow (-{1\over 2}^4)}
and \eqn\lefft{\eqalign{
\wt{|0,k\hskip-3pt>}_{RR}&\rightarrow(\ {1\over 2}^4)\qquad\quad\;
\hbox{For $p$ odd.}\cr
\wt{|0,k\hskip-3pt>}_{RR}&\rightarrow(-{1\over 2}{1\over
2}^3)\qquad \hbox{For $p$ even.}},}
which have opposite chirality.

We have constructed the Ishibashi states for the flat
space-vacuum. We will now construct consistent boundary states.
These states must be invariant under all the symmetries of the
closed string CFT and must factorize via a modular transformation
on the cylinder amplitude. That is, the open string one loop
cylinder amplitude with a given set of boundary conditions must
agree with the answer that results from propagation in the closed
string channel between boundary states
\eqn\factori{
Z(t)=\int{dt\over 2t}\hbox{tr}({1+(-1)^F\over 2}e^{-2tH_o})=\int
dl<\hskip-3pt B| e^{-lH_c}|B\hskip-3pt>,}
such that the open string time $t$ and the closed time $l$ are
related by $t=1/2l$. The boundary states must satisfy the GSO
projection of the underlying string theory they are embedded in.
Therefore, they must satisfy
\eqn\gso{\eqalign{
(-1)^F|B\hskip-3pt>&=(-1)^{\wt F} |B\hskip-3pt>=|B\hskip-3pt>
\qquad \ \ \hbox{Type
IIB}\cr
(-1)^F|B\hskip-3pt>&=-(-1)^{\wt F}|B\hskip-3pt>=|B\hskip-3pt>
\qquad \hbox{Type
IIA}}.}
The GSO operators act on the fermion zero modes (Gamma matrices)
as the chirality
matrix so that $(-1)^F\Gamma^{a\pm}=-\Gamma^{a\pm}(-1)^F$ and
$(-1)^F\psi_r=-\psi_r(-1)^F$ and similarly for the right movers.
Furthermore, since $\vac^{(0)}_{NSNS}$ is odd under both $(-1)^F$ and
$(-1)^{\wt F}$, it follows that the action on the Ishibashi states
\coherent\ is
\eqn\actonishi{
(-1)^F\vac_{NSNS}=-\mvac_{NSNS}\qquad (-1)^{\wt
F}\vac_{NSNS}=-\mvac_{NSNS},}
so that the GSO invariant combination in the NSNS sector is
\eqn\gsoproj{
{1\over \sqrt{2}}(|+\hskip-3pt>_{NSNS}-|-\hskip-3pt>_{NSNS})}
Using \coherent\boundb\lefft\ one gets for the RR sector
\eqn\rrsec{
(-1)^F\vac_{RR}=\mvac_{RR}}
and
\eqn\leftp{
(-1)^{\wt F}\vac_{RR}=\mvac_{RR}\ \hbox{p odd} \qquad(-1)^{\wt
F}\vac_{RR}=-\mvac_{RR}\ \hbox{p even}.}
Therefore, invariance under $(-1)^F$ in the RR sector yields the
following linear combination
\eqn\gsoprojb{
{1\over \sqrt{2}}(|+\hskip-3pt>_{RR}+|-\hskip-3pt>_{RR}).}
Imposing the right moving GSO invariance of \gso\ restricts $p$ to
be odd for the Type IIB superstring and $p$ even for the Type IIA
superstring just as expected for supersymmetric branes\foot{One
can write boundary states for any value of $p$ for the Type
II theories, but if they are not even(odd) for Type IIA(IIB) they
do not carry Ramond-Ramond charge and are unstable due to the
presence of
a tachyon in the open string channel.}.

We must further impose factorizability with the open string
calculation. The open string Hamiltonian is given by
\eqn\hamilo{
H_o=\pi
p^2+\pi\sum_{\mu=0,1,\ldots,7}\left(\sum_{n=1}^{\infty}\am^\mu\ap^\mu
+\sum_{r>0}r\pmi^{\mu}\pp^{\mu}\right)+\pi C_0,}
with $C_0$
zero in the Ramond sector and $-1/2$ in the NS sector. Therefore,
the  partition function for an open string
with $p+1$ Neumann boundary conditions and $9-p$ Dirichlet
boundary conditions in the light cone is
\eqn\parti{\eqalign{
Z=&\int{dt\over 2t}\hbox{tr}({1+(-1)^F\over 2}e^{-2tH_0})=\cr
&={V_{p+1}\over (2\pi)^{p+1}}\left({1\over 2}\right)^{{p+3\over 2}}
\int{dt\over
2t^{{p+3\over 2}}}
{1\over \eta(it)^8}\left(\left({\vt_3(0,it)\over
\eta(it)}\right)^4-\left({\vt_4(0,it)\over
\eta(it)}\right)^4-\left({\vt_2(0,it)\over
\eta(it)}\right)^4\right)}.}
$V_{p+1}$ is the volume of the Dp-brane, the first two terms
correspond to tracing over the NS sector without and with the GSO
insertion $(-1)^F$ and the last term is the trace over the R
sector without any insertion. These can be conveniently written
using $\vt$ functions
\eqn\functions{\eqalign{
\eta(\tau)&=q^{1/24}\prod_{n=1}^\infty(1-q^n)\cr
\vt_1(\nu,\tau)&=2\exp(\pi i \tau/4)\sin(\pi
\nu)\prod_{n=1}^{\infty}(1-q^n)(1-e^{2\pi i \nu}q^n)(1-e^{-2\pi i
\nu}q^n)\cr
\vt_2(\nu,\tau)&=2\exp(\pi i \tau/4)\cos(\pi
\nu)\prod_{n=1}^{\infty}(1-q^n)(1+e^{2\pi i \nu}q^n)(1+e^{-2\pi i
\nu}q^n)\cr
\vt_3(\nu,\tau)&=\prod_{n=1}^{\infty}(1-q^n)
(1+e^{2\pi i \nu}q^{n-1/2})(1+e^{-2\pi i
\nu}q^{n-1/2})\cr
\vt_4(\nu,\tau)&=\prod_{n=1}^{\infty}(1-q^n)
(1-e^{2\pi i \nu}q^{n-1/2})(1-e^{-2\pi i
\nu}q^{n-1/2})},}
where $q=e^{2\pi i \tau}$.

In order to compare with the closed string channel, we must
rewrite \parti\ in closed string time $l=1/2t$. Using modular
properties of $\vt$-functions
\eqn\tranform{\eqalign{
\eta(\tau)&=(-i\tau)^{-1/2}\eta(-1/\tau)\cr
\vt_1(\nu,\tau)&=i(-i\tau)^{-1/2}e^{-\pi i \nu^2/\tau}\vt_1
(\nu/\tau,-1/\tau)\cr
\vt_2(\nu,\tau)&=(-i\tau)^{-1/2}e^{-\pi i \nu^2/\tau}\vt_4
(\nu/\tau,-1/\tau)\cr
\vt_3(\nu,\tau)&=(-i\tau)^{-1/2}e^{-\pi i \nu^2/\tau}\vt_3
(\nu/\tau,-1/\tau)\cr
\vt_4(\nu,\tau)&=(-i\tau)^{-1/2}e^{-\pi i
\nu^2/\tau}\vt_2(\nu/\tau,-1/\tau)}}
 one gets
\eqn\closed{ Z={V_{p+1}\over (2\pi)^{p+1}}{1\over 64}\int{dl\over
l^{{9-p\over 2}}}{1\over
\eta(2il)^8}\left(\left({\vt_3(0,2il)\over
\eta(2il)}\right)^4-\left({\vt_4(0,2il)\over
\eta(2il)}\right)^4-\left({\vt_2(0,2il)\over
\eta(2il)}\right)^4\right).} It is clear that in order to
reproduce this expression using boundary states \factori , that we
must put them in a position eigenstate\foot{Following \sen , we have
chosen to define the corresponding Ramond-Ramond bra vectors without
conjugating the $i$.}
\eqn\bsb{\eqalign{
|\eta\hskip-3pt>_{NSNS}&={\cal N}\int d^{9-p}k\ \vac_{NSNS}\cr
|\eta\hskip-3pt>_{RR}&=4i{\cal N}\int d^{9-p}k\ \vac_{RR}},} so
that the powers of closed string time in \closed\ match.

There is a heuristic way of anticipating which set of "matrix
elements" of  boundary state components result in the various
terms in \closed . We can think of the R sector of the open string
as a twisted sector of the NS sector when we go around $\sigma$.
Moreover, the insertion of $(-1)^F$ in the trace is a twist in the
$t$ direction. Now, since the roles of $t$ and $\sigma$ are
exchanged in the closed string channel, we can make the following
identifications\foot{The normalization constant ${\cal N}$ is
found by requiring these expressions to be equal.}
\eqn\ident{\eqalign{ &\int{dt\over
2t}\hbox{tr}_{NS}(e^{-2tH_0})=\int
{dl}\,_{NSNS}\hskip-3pt<\hskip-3pt
+|e^{-lH_c}|+\hskip-3pt>_{NSNS}= \int
{dl}\,_{NSNS}\hskip-3pt<\hskip-3pt
-|e^{-lH_c}|-\hskip-3pt>_{NSNS}\cr &\int{dt\over
2t}\hbox{tr}_{NS}((-1)^Fe^{-2tH_0})=\int
{dl}\,_{RR}\hskip-3pt<\hskip-3pt +|e^{-lH_c}|+\hskip-3pt>_{RR}=
\int {dl}\,_{RR}\hskip-3pt<\hskip-3pt
-|e^{-lH_c}|-\hskip-3pt>_{RR}\cr &\int{dt\over
2t}\hbox{tr}_{R}(e^{-2tH_0})=\int
{dl}\,_{NSNS}\hskip-3pt<\hskip-3pt
+|e^{-lH_c}|-\hskip-3pt>_{NSNS}= \int
{dl}\,_{NSNS}\hskip-3pt<\hskip-3pt
-|e^{-lH_c}|+\hskip-3pt>_{NSNS}}} and all other "matrix elements"
vanish. One can easily perform the calculation of these matrix
elements using the explicit expressions for the boundary states we
wrote down and using as the closed string Hamiltonian
\eqn\hamiloclo{ H_c=\pi
p^2+2\pi\sum_{\mu=0,1,\ldots,7}\left(\sum_{n=1}^{\infty}\am^\mu\ap^\mu
+\sum_{r>0}r\pmi^{\mu}\pp^{\mu}\right)+2\pi C_0,} with $C_0$ zero
in the RR sector and $-1$ in the NSNS sector.
 One gets the open string
answer if we take our boundary states representing a D-brane to
be\foot{We can easily obtain the expression for an anti-D-brane by
changing the sign of the RR contribution to the boundary state.}
\eqn\fin{ |B\hskip-3pt>={1\over
2}\left(|+\hskip-3pt>_{NSNS}-|-\hskip-3pt>_{NSNS}+|+\hskip-3pt>_{RR}
+|-\hskip-3pt>_{RR}\right).}
and the normalization constant in \bsb\ is ${\cal
N}^2={V_{p+1}\over (2\pi)^{p+1}}{1\over 32}$.

After introducing the necessary ingredients we are now ready to
discuss the realization of fractional branes in orbifold
backgrounds.

\newsec{Fractional Branes and Boundary States}

We will use the open string interpretation of the cylinder
amplitude to classify the consistent set of supersymmetric boundary states
corresponding to D-branes at a ${\bf C}^3/\Gamma$ singularity. We will
analyze
branes that are point-like in the orbifold direction and  in the
transverse space.

An open
string in this background suffers identifications which follow
from the action of $\Gamma$. $\Gamma$ acts both on the coordinates
along the orbifold and on the end-points of the string.
The action on the coordinates is given by the action of $\Gamma$
on the ${\bf 3}$ of $SU(3)$. The
action on the Chan-Paton factors is determined  by choosing a
representation of $\Gamma$ as  in the gauge theory
discussion. Therefore, a general one loop open string amplitude
can be obtained from a set of $n$ cylinder amplitudes obtained by
acting with the $a$-th irreducible representation $\gamma_a$,
$a=1\ldots,n$ on the Chan-Paton factors. Therefore, we expect to be
able to construct as
many basic consistent boundary states of the ${\bf C}^3/\Gamma$ background as
the number of irreducible representations, $n$, that $\Gamma$
admits. The most general boundary state will admit an expansion in
terms of these basic ones.

The $a$-th open string
cylinder amplitude is given by
\eqn\cylindera{
Z^a={1\over|\Gamma|}\sum_{g\in \Gamma}\int{dt\over
2t}\hbox{tr}(g{1+(-1)^F\over 2}e^{-2tH_o}),}
where $H_o$ is the open string Hamiltonian.
This amplitude can be written as
\eqn\decompose{
Z^{a}={1\over |\Gamma|}\sum_{g\in \Gamma}\chi^2_a(g)Z_g(\wt{q}),}
where $\chi_a$ is the character of the $a$-th irreducible
representation of $\Gamma$ and
\eqn\twisted{
Z_g=\int{dt\over
2t}\hbox{tr}(g{1+(-1)^F\over 2}e^{-2tH_o}).}
As mentioned above, these open string amplitudes should have an
interpretation as propagation between boundary states. In this
case proper factorization demands that
\eqn\factorize{
\int dl\,_a\hskip-3pt<\hskip-3pt B | e^{-lH_c}|B\hskip-3pt >_a=Z^a.}
 We will see in different examples how this
identification works and explicitly construct the boundary states
$|B\hskip-3pt>_a$.

Since we are interested in branes that look point-like in the
orbifold direction and in the transverse space, the boundary
conditions that we will impose on the closed string worldsheet fields
are the ones in \bound\ with $p=0$.
A proper linear combination of solutions has to be formed such that the
boundary
states are GSO invariant, $\Gamma$ invariant and  factorize
properly in the open string channel \factorize . The presence of the
orbifold
forces us to consider solutions to \bound\  in all twisted
sectors of the closed string. Therefore,
 the full boundary state associated with the
$a$-th irreducible representation in the open string channel is
formed as a linear combination of boundary states in the $n$
string sectors
\eqn\solu{ |B\hskip-3pt >_a={1\over \sqrt{n}}\sum_{m=0}^{n-1}
|B,m\hskip-3pt
>_a}
We shall see that the matrix elements involving
$|B,m\hskip-3pt>_a$ yield the open string answer when the
corresponding group element is inserted in the open string trace.
 These boundary
states carry charges under twisted sector vertex operators
$|s\hskip-3pt>$ as measured by
the overlap $<\hskip-3pts|B\hskip-3pt>_a$. In particular they carry
charge under
twisted RR fields and as we will see with precisely the correct
value as the corresponding fractional brane in the probe theory
approach. We will now consider several examples including
orbifolds with isolated singularities, non-isolated singularities
and discrete torsion.

\subsec{${\bf C}^3/{{\bf Z}_N}$ Orbifold}

We will first compute the open string cylinder amplitudes that the
boundary states must reproduce.
The action of the ${{\bf Z}_N}$ generator $g$ on the worldsheet fields is
implemented by
\eqn\acton{ g=\exp^{{2\pi i\over N}(a_1s_1+a_2s_2+a_3s_3)},} where
$(s_1,s_2,s_3)$ are vectors in the $SO(6)$ weight lattice
corresponding to  ${\bf C}^3$.
It is therefore useful to introduce three complex coordinates and
their complex conjugate which describe the orbifold
\eqn\complexrepres{\eqalign{
Z^i&={1\over \sqrt{2}}(X^{2i}+i X^{2i+1})\qquad
\bar{Z}^i={1\over
\sqrt{2}}(X^{2i}- iX^{2i+1})\qquad i=1,2,3\cr
\lambda^i&={1\over
\sqrt{2}}(\psi^{2i}+i\psi^{2i+1})\qquad\bar{\lambda}^i={1\over
\sqrt{2}}(\psi^{2i}- i\psi^{2i+1})},}
so that\foot{See appendix for
conventions.}
 \eqn\act{\eqalign{ Z^i&\rightarrow \alpha^{a_i}Z^i\qquad
\b{Z}^i\rightarrow \alpha^{-a_i}\b{Z}^i \qquad a_1+a_2+a_3=0\
(N)\cr \lambda^i&\rightarrow \alpha^{a_i}\lambda^i \qquad
\b{\lambda}^i\rightarrow \alpha^{-a_i}\b{\lambda}^i},} where
$\alpha=e^{2\pi i/N}$ and the action on the open string vacua yield
\eqn\actonvac{\eqalign{ g\cdot|0\hskip-3pt>_{NS}&=|0\hskip-3pt>_{NS}\cr
g\cdot|0\hskip-3pt>_{R}&=16\prod_{i=1}^{3}\cos(\pi\nu_i)
|0\hskip-3pt>_{R}},} where
$\nu_i=a_i/N$.

Performing the trace in \twisted\ yields
\eqn\unt{
Z_{g^0}={L\over 2\pi}{1\over 2\sqrt{2}}\int{dt\over
2t^{3/2}}{1\over \eta(it)^8}\left(\left({\vt_3(0,it)\over
\eta(it)}\right)^4-\left({\vt_4(0,it)\over
\eta(it)}\right)^4-\left({\vt_2(0,it)\over
\eta(it)}\right)^4\right),}
and
\eqn\parti{\eqalign{
Z_{g^m}\hskip-3pt={L\over 2\pi}{4\over \sqrt{2}}\int \hskip-4pt
{dt\over
2t^{3/2}}{1\over \eta(it)^3}\prod_{i=1}^3{\sin(\pi m \nu_i)\over
\vt_1(m\nu_i,it)}\Biggl(\hskip-3pt&\vt_3(0,it)\hskip-3pt
\prod_{i=1}^3\hskip-3pt\vt_3(m\nu_i,it)-
\vt_4(0,it)\hskip-3pt\prod_{i=1}^3\hskip-3pt\vt_4(m\nu_i,it)\cr
&-\vt_2(0,it)\hskip-3pt\prod_{i=1}^3\hskip-3pt\vt_2(m\nu_i,it)\Biggl),}}
where $m=1,
\ldots,N-1$ and $L/2\pi$ comes from integration over $x^0$.
 Worldsheet duality relates
this open string amplitude to the propagation in closed string
time $l=1/2t$  between boundary states. Therefore, we must rewrite
\parti\ in closed string time using modular properties of $\vt$ functions.
 This yields the following
expressions
\eqn\zero{
Z_{g^0}={L\over 2\pi}{1\over 64}\int{dl\over
l^{9/2}}{1\over \eta(2il)^8}\left(\left({\vt_3(0,2il)\over
\eta(2il)}\right)^4-\left({\vt_4(0,2il)\over
\eta(2il)}\right)^4-\left({\vt_2(0,2il)\over
\eta(2il)}\right)^4\right),}
and
\eqn\partitrans{\eqalign{
Z_{g^m}=i{L\over 2\pi}\int &{dl\over
l^{3/2}}{1\over \eta(2il)^3}\prod_{i=1}^3{\sin(\pi m \nu_i)\over
\vt_1(-2im\nu_il,2il)}\Biggl(\vt_3(0,2il)\prod_{i=1}^3\vt_3
(-2im\nu_il,2il)\cr
&-
\vt_4(0,2il)\prod_{i=1}^3\vt_4(-2im\nu_il,2il)
-\vt_2(0,2il)\prod_{i=1}^3\vt_2(-2im\nu_il,2il)\Biggl)}.}
The expressions we will find next for the boundary states will
reproduce this one and will thus constitute consistent
D-brane states.

\subsec{Boundary States in ${\bf C}^3/{{\bf Z}_N}$}

We will now show that the term in the open string partition
function with $g^m$ inserted in the trace \partitrans\ is reproduced by
taking the solutions of \bound\ in the $m$-th twisted sector of
the closed string and constructing the corresponding coherent
states in the RR and NSNS sector.

The untwisted sector solutions to \bound\ are identical to those
in flat space just discussed. However, we must also make sure that the
boundary state we write down is ${{\bf Z}_N}$ invariant. It is therefore
convenient to write the Ishibashi states using the complex
coordinates \act . Then\foot{We introduce another label for the
boundary states which
describes the sector of closed string oscillators we use.},
\eqn\formcompl{\eqalign{
|\eta,k,0\hskip-3pt>_{NSNS \atop RR}=\exp\Biggl(\sum_{n=1}^{\infty}{1\over
n}(-\am^0\wt\am^0+\am^1\wt\am^1+\sum_{i=1}^3
\beta_{-n}^i\wt{\bar{\beta}}^i_{-n}+\bar{\beta}_{-n}^i\wt\beta^i_{-n})\cr
+i\eta\sum_{r>0}(-\psi^0_{-r}\wt\psi^0_{-r}+\psi^1_{-r}\wt\psi^1_{-r}+
\sum_{i=1}^3\lambda^i_{-r}\wt{\bar
\lambda}^i_{-r}+\bar{\lambda}^i_{-r}\wt{
\lambda}^i_{-r})\Biggl)|\eta,k,0\hskip-3pt>^{(0)}_{NSNS \atop RR}}.}
 $\beta,\bar{\beta},\lambda,\bar{\lambda}$ are the left moving
 oscillators of the complex worldsheet fields and similarly for
 the right movers\foot{See appendix for mode expansions.}.
Written in this form it is clear that the exponential is neutral
under the action of ${{\bf Z}_N}$ since each oscillator appears
multiplying a complex conjugate one. Furthermore, using \acton\
and the explicit expressions for the vacua we see that in fact the
full set of Ishibashi states \formcompl\ are ${{\bf Z}_N}$ invariant in the
untwisted
sector. Therefore the expression reproducing the open string
result without any ${{\bf Z}_N}$ element in the trace \zero\ is \fin\ with
\eqn\solutione{
{\cal N}^{a2}_0={L\over  2\pi}{1\over 32}.}

In closed string theory the string can be closed up to an action
of $g^m$. This sector of the string is usually referred to the
$m$-twisted sector. It is easy to see that in a twisted sector the
modding of the oscillators along the orbifold directions get
shifted. In the ${\bf C}^3/{{\bf Z}_N}$ example it is easy to see from the mode
expansions
in the Appendix that the
modified moddings in the $m$-th twisted sector are
\foot{Since we are interested in isolated singularities we
consider $N$ odd. We will consider examples of non-isolated
singularities in an upcoming section.} \eqn\modd{\eqalign{
\beta^i_{n+m\nu_i}\qquad \wt\beta^i_{n-m\nu_i}\qquad
\bar{\beta}^i_{n-m\nu_i}\qquad \wt{\bar \beta}^i_{n+m\nu_i}\cr
\lambda^i_{r+m\nu_i}\qquad \wt\lambda^i_{r-m\nu_i}\qquad
\bar{\lambda}^i_{r-m\nu_i}\qquad \wt{\bar \lambda}^i_{r+m\nu_i}}}
which satisfy \eqn\commutacokp{\eqalign{
[\beta_{n+m\nu_i}^i,\bar{\beta}_{p-m\nu_i}^j]&=
[\wt\beta_{n-m\nu_i}^i,\wt{\bar\alpha}_{p+m\nu_i}^j]
=(n+m\nu_i)\delta_{n+p}\delta_{ij}\cr
\{\lambda^i_{r+m\nu_i},\bar{\lambda}^j_{s-m\nu_i}\}&=
\{\wt\lambda^i_{r-m\nu_i},
\wt{\bar{\lambda}}^j_{s+m\nu_i}\}=\delta_{r+s} \delta_{ij}}} with
the rest of (anti)commutators vanishing. Moreover, the orbifold
action projects out the bosonic and fermionic zero modes
 along the
orbifold directions since twisted sector states are stuck at the
origin of the singularity. Technically, we have to perform the
same construction as in flat space. The boundary conditions to be
solved are those in \bound\ with the shifted oscillators in the
orbifold directions \eqn\modeqn{\eqalign{
(\beta^i_{n+m\nu_i}-\wt\beta^i_{-n-m\nu_i})|\eta,k,m\hskip-3pt>_{NSNS\atop
RR}&=0\cr
(\bar{\beta}^i_{n-m\nu_i}-\wt{\bar{\beta}}^i_{-n+m\nu_i})
|\eta,k,m\hskip-3pt>_{NSNS\atop RR}&=0\cr
(\lambda^i_{r+m\nu_i}-i\eta\wt\lambda^i_{-r-m\nu_i})
|\eta,k,m\hskip-3pt>_{NSNS\atop RR}&=0\cr
(\bar{\lambda}^i_{r-m\nu_i}-i\eta\wt{\bar{\lambda}}^i_{-r+m\nu_i})
|\eta,k,m\hskip-3pt>_{NSNS\atop RR}&=0\cr}.} Therefore, the
Ishibashi states in the $m$-th twisted sector are given by
\eqn\ishi{\eqalign{ |\eta,k,m\hskip-3pt>_{NSNS \atop
RR}=&\exp\Biggl(\sum_{n=1}^{\infty} {1\over
n}(-\am^0\wt\am^0+\am^1\wt\am^1)+\sum_{i=1}^3 {1\over
n-m\nu_i}\beta_{-n+m\nu_i}^i\wt{\bar{\beta}}^i_{-n+m\nu_i}\cr+
&{1\over n+m\nu_i}\bar{\beta}_{-n-m\nu_i}^i\wt\beta^i_{-n-m\nu_i}
+i\eta\sum_{r>0}(-\psi^0_{-r}\wt\psi^0_{-r}+\psi^1_{-r}\wt\psi^1_{-r}\cr
&+ \sum_{i=1}^3\lambda^i_{-r+m\nu_i}\wt{\bar
\lambda}^i_{-r+m\nu_i}+\bar{\lambda}^i_{-r-m\nu_i}\wt{
\lambda}^i_{-r-m\nu_i})\Biggl)|\eta,k,m\hskip-3pt>^{(0)}_{NSNS
\atop RR}},} where $|\eta,k,m\hskip-3pt>^{(0)}_{RR}$  is
constructed as before but only with the zero modes of the
transverse coordinates $x^0,x^1$. Therefore, the GSO, ${{\bf
Z}_N}$ invariant boundary state in the $m$-th twisted sector
is\foot{The GSO operator acts the same as usual on oscillators and
as the chirality matrix on the zero modes.}
 \eqn\twistedsect{ |B,m\hskip-3pt>^a={1\over
2}\left(|+,m\hskip-3pt>^a_{NSNS}-|-,m\hskip-3pt>^a_{NSNS}+
|+,m\hskip-3pt>^a_{RR}+|-,m\hskip-3pt>^a_{RR} \right)\qquad
m=1,\ldots,N-1.}
 However, since the orbifold removes the zero modes
along the orbifold, now the Fourier transform is only in the
transverse directions
\eqn\bsba{\eqalign{
|\eta,m\hskip-3pt>^a_{NSNS}&={\cal N}^a_m\int d^{3}k\
|\eta,m,k\hskip-3pt>_{NSNS}\cr
|\eta,m\hskip-3pt>_{RR}&=\sqrt{2}i{\cal N}^a_m\int d^{3}k\
|\eta,m,k\hskip-3pt>_{RR}}.}
Using the explicit form of these boundary states, we will now see
that the different matrix elements in \twistedsect\ yield the open
string answer \partitrans. Therefore,
\eqn\ident{\eqalign{
\int
{dl}\,_{NSNS}&\hskip-3pt^a\hskip-3pt<\hskip-3pt +,m|e^{-lH_c}
|+,m\hskip-3pt>^a_{NSNS}=
\int
{dl}\,_{NSNS}\hskip-3pt^a\hskip-3pt<\hskip-3pt
-,m|e^{-lH_c}|-,m\hskip-3pt>^a_{NSNS}=\cr
&=i{\cal
N}_m^{a2}\int
{dl}{1\over l^{3/2}}\prod_{i=1}^3{1\over
\vt_1(-2im\nu_il,2il)}\vt_3(0,2il)\prod_{i=1}^3\vt_3(-2im\nu_il,2il)\cr
\int
{dl}\,_{RR}&\hskip-3pt^a\hskip-3pt<\hskip-3pt +,m|e^{-lH_c}
|+,m\hskip-3pt>^a_{RR}=
\int
{dl}\,_{RR}\hskip-3pt^a\hskip-3pt<\hskip-3pt
-,m|e^{-lH_c}|-,m\hskip-3pt>^a_{RR}=\cr
&=-i{\cal
N}_m^{a2}\int
{dl}{1\over l^{3/2}}\prod_{i=1}^3{1\over
\vt_1(-2im\nu_il,2il)}\vt_4(0,2il)\prod_{i=1}^3\vt_4(-2im\nu_il,2il)\cr
\int
{dl}\,_{NSNS}&\hskip-3pt^a\hskip-3pt<\hskip-3pt +,m|e^{-lH_c}
|-,m\hskip-3pt>^a_{NSNS}=
\int
{dl}\,_{NSNS}\hskip-3pt^a\hskip-3pt<\hskip-3pt
-,m|e^{-lH_c}|+,m\hskip-3pt>^a_{NSNS}=\cr
&=i{\cal
N}_m^{a2}\int
{dl}{1\over l^{3/2}}\prod_{i=1}^3{1\over
\vt_1(-2im\nu_il,2il)}\vt_2(0,2il)\prod_{i=1}^3\vt_2(-2im\nu_il,2il)}}
and all other "matrix elements" vanish.

Thus, the $m$-twisted sector contribution to the boundary state
reproduces the open string amplitude with $g^m$ inserted in the
trace if we choose
\eqn\constants{
{\cal N}^{a2}_m={L\over 2\pi}{\chi^{a2}(g^m)\over 4}8\prod_{i=1}^3
\sin(\pi m \nu_i).}
We have therefore constructed a consistent boundary state for
each irreducible representation of ${{\bf Z}_N}$, as expected. The factors
of $\sin$ in \constants\ might seem awkward, but in fact they are
expected. Even though the orbifold action leaves fixed only the origin,
in order to get a modular invariant closed string partition
function, we must multiply the contribution from a given twisted
sector by the fixed point degeneracy that the orbifold would have
if it were toroidal. Now, for toroidal orbifolds, the number of
fixed of the group element $g^m$ is given by Lefschetz
fixed-point theorem
\eqn\lefts{
\det(1-g^m)=64\prod_{i=1}^3\sin^2(\pi m \nu_i),}
as can be shown using the defining action \acton , and \constants\
contains the square root of this.

The charges carried by the boundary state can be computed by
inserting the corresponding vertex operator. Thus,
the charge vector of these states is made out of untwisted
D0-brane charge and the twisted RR charges. The state that
represents a D0-brane is the one made with the
regular representation. Since the character of the regular
representation is non-trivial only for the identity element, it is
clear that the corresponding boundary state is  charged only under
the untwisted RR one-form as expected from the probe theory
considerations.
The D0-brane charge of the above states
built on an irreducible representation is
\eqn\dnotcharge{
Q^a_0={1/N}\qquad a=1,\dots,N.}
These boundary states also carry charges under the RR twisted
sectors. The charge under the $m$-th such sector can be easily
computed to be
\eqn\twistedcharge{
Q_m^a={\chi^a(g^m)\over N}\qquad a=1,\dots,N,}
as expected from the disk computation.

We have constructed all the boundary states corresponding to
fractional branes for this orbifold. Any other supersymmetric
D-brane state can
be constructed out of these basic ones. These states can be
identified in the large volume limit of the orbifold CFT moduli
space with D-branes wrapping supersymmetric cycles.

\subsec{${\bf C}^3/{{\bf Z}_N}\times {{\bf Z}_N}$ Orbifold}

The orbifold group is generated by two elements $g_1$ and $g_2$,
so that an arbitrary group element $g_1^pg_2^q$ acts as
\eqn\act{\eqalign{
Z^1&\rightarrow \alpha^{p}Z^1\qquad Z^2\rightarrow
\alpha^{q}Z^2 \qquad Z^3\rightarrow
\alpha^{-(p+q)}Z^3\quad p,q=0,\ldots,N-1\cr
\psi^1&\rightarrow
\alpha^{p}\psi^1 \qquad \psi^2\rightarrow
\alpha^{q}\psi^2 \qquad \psi^3\rightarrow
\alpha^{-(p+q)}\psi^3},}
where $\alpha=e^{2\pi i/N}$ and the action on the vacua is
\eqn\actonvac{\eqalign{
g\cdot|0>_{NS}&=|0>_{NS}\cr
g\cdot|0>_{R}&=16\cos({\pi p\over N})\cos({\pi q\over N})
\cos({\pi (p+q)\over N})|0>_{R}},}
This is an isolated singularity.
There are $(N-1)(N-2)$ elements of
this group that fix the origin. The contribution to the partition
function with these group elements  or the identity in the trace
\twisted\ is identical to that in the
previous example \unt\parti\ with the substitutions
$k\nu_1\rightarrow p$, $k\nu_2\rightarrow
q$ and $k\nu_3\rightarrow -(p+q)$. Therefore the modular transform
is \zero\partitrans .

 There are $3(N-1)$ group
elements that fix a line. The contribution of these group elements
to the open string partition function can be accounted for by
adding three copies of
\eqn\contrifix{\eqalign{
 Z_{g}(it)={L\over 2\pi}{2\over
\sqrt{2}}\int{dt\over 2t^{3/2}}{1\over \eta(it)^6}&{\sin({\pi
p\over N})^2\over
\vt_1(p/N,it)^2}\Biggl(\vt_3(0,it)^2\vt_3(p/N,it)^2\cr
&-
\vt_4(0,it)^2\vt_4(p/N,it)^2-\vt_2(0,it)^2\vt_2(p/N,it)^2
\Biggl)}}
where $p=1,\ldots
N-1$. The modular transform of this expression that we must
reproduce from boundary states is
\eqn\sdualnoniso{\eqalign{
 Z_{g}(2il)=-{L\over 2\pi}{1\over
4}\int{dl\over l^{5/2}}&{1\over \eta(2il)^6}{\sin({\pi
p\over N})^2\over
\vt_1(-2ipl/N,2il)^2}\Biggl(\vt_3(0,2il)^2\vt_3(-2ipl/N,2il)^2\cr
&-
\vt_4(0,2il)^2\vt_4(-2ipl/N,2il)^2-\vt_2(0,2il)^2\vt_2(-2ipl/N,2il)^2
\Biggl)}}

\subsec{Boundary States in ${\bf C}^3/{{\bf Z}_N}\times {{\bf Z}_N}$}

The boundary state is again a sum over all twisted sectors
\eqn\sumover{
|B\hskip-3pt >_a={1\over N}\sum_{m=0}^{N^2-1}|B,m\hskip-3pt>_a.}
The sectors associated with group elements fixing a point can be
obtained from the previous example \twistedsect,\ \constants\
with the above mentioned
substitutions. The sectors corresponding to group elements fixing
a line can be constructed as if we were dealing with a
${\bf C}^2/{{\bf Z}_N}$
orbifold singularity. There are extra zero modes for these so that
\eqn\bsbab{\eqalign{
|\eta,m\hskip-3pt>^a_{NSNS}&={\cal N}^a_m\int d^{5}k\
|\eta,m,k\hskip-3pt>_{NSNS}\cr
|\eta,m\hskip-3pt>_{RR}&=2i{\cal N}^a_m\int d^{5}k\
|\eta,m,k\hskip-3pt>_{RR}},}
with the obvious modifications in \ishi . Factorization with the
open string requires
\eqn\normali{
{\cal N}^a_g={L\over 2\pi}{\chi^{a2}(g)\over 8}4\sin^2({\pi p\over N}),}
where the fixed point degeneracy appears as expected.
In this example the untwisted D0-brane charge is
\eqn\carrega{
Q^a_0={1\over N^2}\qquad a=1,\dots,N^2}
and the charge under the $m$-th twisted RR field
\eqn\carregatwist{
Q^a_m={\chi^a(g^m)\over N^2}\qquad a=1,\dots,N^2.}

\subsec{${\bf C}^3/{{\bf Z}_N}\times {{\bf Z}_N}$ With Discrete Torsion}

It is easy to construct the boundary states describing fractional
branes for this vacuum. Technically, the only difference is the
use of projective representations, and therefore the appearance of
modified characters in the open string computation. Once the
discrete torsion cocycles are chosen -- $H^2({\bf Z}_N\times {\bf
Z}_N,U(1))\simeq {\bf Z}_N$, so that there are $N-1$ nontrivial
cohomology classes -- we can construct exactly as
in the previous example boundary states corresponding to 
fractional branes.
However, their number is
the number of irreducible representations in the cocycle class of
the discrete torsion phases. Within each cocycle class there is a
unique $N$ dimensional irreducible projective representation so that 
there is a unique
fractional brane. Its charges are those in \carrega,\ \carregatwist\
for $a=1$  with the modified character of the projective
representation.

\newsec{Geometric Interpretation}

The construction of BPS D-particle states has been so far restricted
to exactly solvable orbifold theories, where the boundary state approach
proved to be a powerful tool. However, orbifold theories admit exactly
marginal deformations which give rise to a moduli space. This fact
leads to a natural question, namely how does the BPS spectrum behave
under these perturbations? Although a precise answer may depend on the
particular aspects of each model, we will outline the general expected
behavior and establish connections to singularity theory and McKay
%\nref\husong{Y-H. He, J.S. Song, "Of McKay Correspondence, 
%Non-linear Sigma-model and Conformal Field
%Theory", hep-th/9903056.}%  
correspondence.
In the next section, we  will treat the
case of ${\bf C}^3/{{\bf Z}_3}$ in detail, as in illustration of the
principles outlined below. Since the effect of marginal deformations
is very different in theories with and without discrete torsion,
we start the discussion with conventional orbifolds. The case
with discrete torsion will be considered at the end of the section.

In the absence of discrete torsion, the marginal operators correspond to
blow-up modes of the singularity, parameterizing the \ka\ moduli space
of the resolved space
\A .
Moreover, by adjusting the coefficients of the perturbations,
we can eventually reach a region in the moduli space where the
resulting exceptional cycles are very large and classical geometry
is a good approximate description of the theory. Assuming that the
BPS spectrum can be continuously deformed to this region
with no jumping phenomena, we expect the orbifold states to
be realized as D-branes wrapped on supersymmetric cycles
in a smooth space-time background. Note that in this regime, there is
no exact conformal field theory description, but the supergravity
approximation is valid. Therefore we would obtain two different
realizations of the same spectrum of BPS states. The
assumption that jumping phenomena are absent is automatically
satisfied in space-time theories with sixteen supercharges.
In theories with eight supercharges, where this phenomenon is present,
we simply assume that there is a path connecting the orbifold point
to the large radius limit that avoids the curves of marginal stability.

For isolated singularities, carrying out this program is in fact
equivalent to a physical realization of the celebrated McKay
correspondence
\refs{\McK-\IN}, as we now explain.
Loosely, we can think of supersymmetric
states in the large radius limit as D-branes wrapping supersymmetric
cycles in space-time. The homology of the resolved space is even and
generated by holomorphic cycles, which are supersymmetric. At the same
time, the orbifold boundary states are in one-to-one correspondence 
with the
irreducible representations of the orbifold group.
Therefore, the deformation of BPS states effectively gives a map
between the irreducible representations of the orbifold group and
homology classes of the resolution. This would be a physical realization
of the McKay correspondence.

A more precise formulation\foot{This paragraph is somewhat
abstract and it is  not essential for understanding the rest of the
paper.}
can be achieved by
interpreting the D0-brane states as K-theory classes
\nref\MM{R. Minasian and G. Moore, ``K-theory and Ramond-Ramond charge'',
JHEP {\bf 11} (1997) 002, hep-th/9710230.}%
\refs{\MM,\W}.
In the present context, D-brane
states in an orbifold $X/\Gamma$ are classified by the equivariant
K-theory $K_\Gamma(X)$ with compact support of the covering space
\nref\PH{P. Horava, ``Type IIA D-Branes, K-Theory, and Matrix Theory'',
Adv. Theor. Math. Phys. {\bf 2} (1998) 1373, hep-th/9812135.}%
\nref\GC{H. Garcia-Compean, ``D-Branes in Orbifold Singularities and
Equivariant K Theory'', hep-th/9812226.}%
\nref\SG{S. Gukov, ``K Theory, Reality and Orientifolds'',
hep-th/9901042.}%
\nref\BGP{O. Bergman, E.G. Gimon and P. Horava, 
``Brane Transfer Operations and T-Duality of Non-BPS States'',
JHEP {\bf 04} (1999) 010, hep-th/9902160.}%
\refs{\W,\PH,\GC,\SG,\BGP}.
A standard result \ref\Seg{G. Segal, ``Equivariant K-theory'',
Institut Des Hautes \'Etudes Scientifiques,
Publications Math\'ematiques {\bf 34} (1968) 129.}
shows that $K_\Gamma(X)$ is isomorphic to the representation group
$R(\Gamma)$ of the discrete group $\Gamma$ generated by its 
irreducible representations. As mentioned before, this is
in agreement with the boundary state construction.
In the large radius limit,
the D-brane states are classified similarly by a K-theory group
supported on the exceptional divisor $D$ of the resolved space
$\widetilde{X}$. More precisely, the relevant K group
$K_0({\widetilde{X}})$ can be defined
as the Grothendieck group of bounded complexes of algebraic
vector bundles
supported on the exceptional locus\foot{This construction can be
realized both in algebraic and topological setting, resulting in general
in different objects. Since we are ultimately interested in BPS
configurations, we will consider here the algebraic approach.}
\nref\BFM{P. Baum, W. Fulton and R. MacPherson,
``Riemann-Roch and Topological K-theory for Singular Varieties'',
Acta Mathematica {\bf 143} (1979) 155.}%
\refs{\Seg,\BFM}. The relevance of this K-theory group
in the context of D-branes on orbifolds and more general algebraic
varieties has been discussed in
\nref\ES{E. Sharpe, ``D-Branes, Derived Categories, and
Grothendieck Groups'',\hfill\break hep-th/9902116.}%
\refs{\SG,\ES}. It can also be showed that $K_0({\widetilde{X}})$
is isomorphic to the usual Grothendieck group $K(D)$ of coherent sheaves
of the exceptional divisor $D$. When $D$ is smooth, $K(D)$ is generated
by classes of vector bundles. In this framework, the deformation
of BPS states will give a map between the K-theory groups
$K_\Gamma(X)$ and $K_0({\widetilde{X}})$. This is a more precise
formulation of the McKay correspondence \refs{\GSV,\AV,\IN}.

In practice, deforming the spectrum of BPS
states along the moduli space can be quite difficult,  especially
in theories with eight supercharges which exhibit quantum
corrections. The ${\bf C}^3/{{\bf Z}_3}$ example studied in the next
section illustrates the complexity of the problem.

For technical reasons, the above discussion has been restricted to
isolated singularities. However, as detailed in section 3.3, the
boundary state construction works as well for nonisolated
singularities. We have showed that the fractional branes are again
classified by the irreducible representations of the orbifold
group. Therefore, in the ${\bf C}^{3}/ {{\bf Z}_N}\times {{\bf
Z}_N}$ orbifold treated there, we have $(N^2-1)$ independent
boundary states. The number of homology cycles of the exceptional
locus can be computed using either conformal field theory
techniques or toric methods. The abstract CFT Hodge diamond is
\DF\
\eqn\hodge{\matrix {& & & &0& & &  \cr
           & & &0& &0& &  \cr
           & &0& &{(N+4)(N-1)\over 2}& &0&  \cr
           &0& &0& &0& &0 \cr
           & &0& &{(N+4)(N-1)\over 2}& &0&  \cr
           & & &0& &0& &  \cr
           & & & &0& & & }}
Therefore, there seems to be a mismatch with the number of irreducible
representations. A more careful analysis\foot{In the computation of the
orbifold cohomology performed in the section 2.2 of \DF\ one has to
remove the $3(N-1)$ twisted sector $(1,1)$ classes corresponding to
the constant forms of zero degree along the fixed lines. These give
rise to non-normalizable
forms on the resolved space.} shows that the Hodge diamond
corresponding to normalizable cohomology classes or,
by Poincar\'e duality, to compact cycles, is
\eqn\hodge{\matrix {& & & &0& & &  \cr
           & & &0& &0& &  \cr
           & &0& &{(N-2)(N-1)\over 2}& &0&  \cr
           &0& &0& &0& &0 \cr
           & &0& &{(N+4)(N-1)\over 2}& &0&  \cr
           & & &0& &0& &  \cr
           & & & &0& & &   }}
which is in agreement with the number of fractional branes.
The same result can be obtained by constructing toric resolutions.
This suggests that the map between representations of the orbifold
group and cycles of the resolution can be extended to this case,
at least in the D-brane picture. It would be interesting to
give a more precise description of this map, but we leave this for
future work.

Finally, the case of orbifolds with discrete torsion is notably
different. The exactly marginal operators correspond to complex
structure deformations, as opposed to \ka\ blow-up modes. Moreover,
turning on these operators does not completely resolve the singularity
\nref\AM{P. Aspinwall and D. Morrison, ``Stable Singularities in String
Theory'', Commun. Math. Phys. {\bf 178} (1996) 115, hep-th/9503208.}%
\refs{\VW,\AM,\Da,\DF}. Therefore, the fractional branes cannot be
given a clear geometric interpretation.
We believe that these states can be mathematically
described
in an appropriate K-theoretic formalism, but we will not attempt
to develop this here.

\newsec{Fractional Branes in The ${{\bf C}^3}/{{\bf Z}_3}$ Orbifold}

The purpose of this section is to carry out the program outlined above
for the ${\bf C}^3/{\bf Z}_3$ orbifold.
The moduli space of deformations can be
thought as the \ka\ moduli space of a noncompact Calabi-Yau threefold
with a ${\bf P}^2$ cycle shrinking to zero size
\nref\AGM{P.S. Aspinwall, B.R. Greene and D.R. Morrison, ``Measuring
Small Distances in $N=2$ Sigma Models'', \np{420}{1994}{184},
hep-th/9311042.}%
\refs{\A,\AGM}. As usual in $(2,2)$
superconformal models, the classical geometry is corrected by
world-sheet instantons whose effects can be exactly summed using
local mirror symmetry. Therefore, the first step in our  analysis
is to describe in detail the exact quantum moduli space.
Then, using these exact results, we show how to continuously
deform the fractional branes to the large radius limit.

\subsec{The Quantum Moduli Space}

Since this is a nonconventional case --the threefold
being noncompact-- we have in fact to use the local version of
mirror symmetry developed in
\refs{\A, \KKV,\KMV,\CKYZ}.
The coefficients of the marginal deformations
can be thought as algebraic coordinates on the moduli space and
the exact quantum geometry is described by periods of the local mirror
geometry. The present model has been analyzed to various degrees in
\nref\KMVb{A. Klemm, P. Mayr and C. Vafa, ``BPS States of Exceptional
Non-Critical Strings'', Nucl. Phys. {\bf B} (Proc. Suppl.) {\bf 58}
(1997) 177, hep-th/9607139.}%
\nref\LMW{W. Lerche, P. Mayr and N. Warner, ``Non-Critical Strings,
Del Pezzo Singularities and Seiberg-Witten Curves'', \np{499}{1997}{125},
hep-th/9612085.}%
\nref\KZ{A. Klemm and E. Zaslow, ``Local Mirror Symmetry at
Higher Genus'', hep-th/9906046}%
\refs{\A,\AGM,\KMVb,\LMW,\CKYZ,\KZ}.
In the following we will review the construction of the local mirror
model and present a detailed solution.
The results in this subsection have been elaborated in collaboration
with Albrecht Klemm to whom we are very grateful for valuable help.

The starting point is the linear sigma model construction of the
(blown-up) orbifold background in IIA string theory. Following
\nref\W{E. Witten, ``Phases of $N=2$ Theories in Two Dimensions'',
\np{403}{1993}{159}, hep-th/9301042.}%
\refs{\W,\DGM}, we consider a two dimensional $N=2$ $U(1)$ sigma model
with four chiral fields $X_i$, $i=0,\ldots,3$ and charge vector
\eqn\chvect{
l=(-3,1,1,1).}
The potential for the scalar fields is therefore
\eqn\potential{
V=\left(|X_1|^2+|X_2|^2+|X_3|^2-3|X_0|^2-r\right)^2}
where $r$ is a Fayet-Iliopoulos parameter. The phase $r>0$ corresponds
to the blown-up phase, the exceptional $P^2$ divisor being described
by the coordinates $X_1,X_2,X_3$ which cannot vanish simultaneously.
In the phase $r<0$, $X_0$ cannot vanish and the exceptional divisor
is blown down. It will be shown in the following that this picture is
corrected quantum mechanically.
In the blown-up phase, the moduli space of the linear sigma model
is a noncompact toric variety described by the following noncomplete
fan

\ifig\toricfig{The trace of the noncomplete fan corresponding to the
blown-up ${{\bf C}^3}/{{\bf Z}_3}$. The black node in the center
represents the
compact exceptional divisor $D\simeq {\bf P}^2$.}
{\epsfxsize1.5in\epsfbox{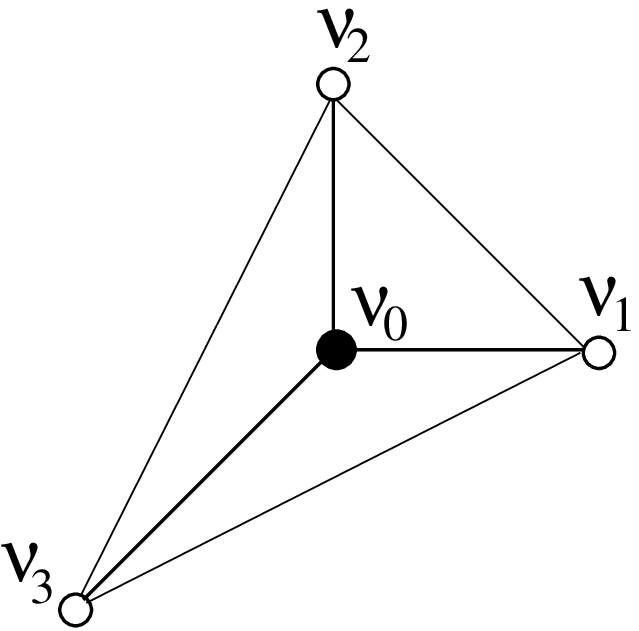}}

\noindent
The associated toric data are
\eqn\toric{
\nu_0=\left[\matrix{0\cr 0\cr 1\cr}\right],\qquad
\nu_1=\left[\matrix{1\cr 0\cr 1\cr}\right],\qquad
\nu_2=\left[\matrix{0\cr 1\cr 1\cr}\right],\qquad
\nu_3=\left[\matrix{-1\cr -1\cr 1\cr}\right].}
Local mirror symmetry associates to the noncompact toric variety
represented in Fig.1. a one dimensional local geometry described by the
polynomial equation \refs{\KMV,\CKYZ}
\eqn\pol{
\sum_{i=0}^3 a_iy_i=0}
where the variables $y_i$ satisfy the constraint equation
\eqn\cosntr{
y_1y_2y_3y_0^{-3}=1.}
The solution can be easily presented in parametric form
\eqn\sol{
y_1=x_1^3,\qquad y_2=x_2^3,\qquad y_3=x_3^3,\qquad y_0=x_1x_2x_3}
where $x_1,x_2,x_3$ are projective coordinates subject to the
${\bf C}^*$ identification $(x_1,x_2,x_3)\sim (\mu x_1,\mu x_2,\mu x_3)$.
Note that there is also an $\left({\bf C}^*\right)^3$ action on the
space of parameters $a_i$
\eqn\scaling{
(a_0,a_1,a_2,a_3)\rightarrow (\alpha_1\alpha_2\alpha_3a_0, \alpha_1^3a_1,
\alpha_2^3a_2,\alpha_3^3a_3)}
which leaves equation \pol\ invariant. Therefore the moduli space is
one dimensional and can be parameterized by the invariant coordinate
\eqn\inv{
z=-27{a_1a_2a_3\over a_0^3}}
where the factor $-27$ has been introduced for later convenience.
The local mirror geometry is then described by the elliptic
fibration
\eqn\ellfiber{
x_1^3+x_2^3+x_4^3-\psi x_1x_2x_3=0,}
with $z={27\over \psi^3}$,
which has been considered in a different context before
\nref\KLRY{A. Klemm, B.H. Lian, S.S. Roan and S.T. Yau, ``A Note
on ODEs from Mirror Symmetry'',  hep-th/9407192.}%
\nref\LY{B.H Lian and S.T. Yau, ``Arithmetic Properties of Mirror Map
and Quantum Coupling'', \cmp{176}{1996}{163}, hep-th/9411234.}%
\refs{\KLRY,\LY}.

The exact solution of the model is provided by a three-dimensional
vector of periods satisfying the differential equation\foot{Note that
this equation is derived by applying the general mirror construction to
this local case. This is not the Picard-Fuchs equation for the periods
of the elliptic curve \ellfiber\ which will enter the discussion later.}
\eqn\PF{
\left[\theta_z^3-z\left(\theta_z+{1\over 3}\right)
\left(\theta_z+{2\over 3}\right)\theta_z\right]f=0}
where $\theta_z=z{d\over dz}$.
This is a particular case of a general class of differential
equations defining the Meijer G-functions $G^{m,n}_{p,q}
\left(z\left|\matrix{&\hfill a_1\ldots a_p\hfill\cr
&\hfill b_1\ldots b_q\hfill\cr}
\right.\right)$
\ref\Mj{C.S. Meijer, ``On the G-function I'', Indagationes
Mathematicae Vol {\bf XLIX}, (1946), 124; ``On the G-function II'',
Indagationes
Mathematicae Vol {\bf XLIX}, (1946), 213.}
\eqn\Mjeq{
\left[(-1)^{p-m-n}\prod_{j=1}^p(\theta_z-a_j+1)-
\prod_{j=1}^q(\theta_z-b_j)
\right]f=0}
where $0\leq n\leq p\leq q$ and $0\leq m\leq q$.
Note that after a change of variables $z\rightarrow -z$, the equation
\PF,\ becomes a Meijer equation with $p=q=3$ and
\eqn\param{
a_1={1\over 3},\qquad a_2={2\over 3},\qquad a_3=1, \qquad b_1=b_2=b_3=0.}
This shows that \PF\ has regular singular points at $z=0,1,\infty$.
According to the general theory, first we have to find local solutions
defined in a neighborhood of each singular point and then
perform analytic continuation. The local solutions will be denoted
by $f^x_i(z)$ where $x=0,1,\infty$ labels the singular points and
$i=0,1,2$ labels solutions in a given region. Note that the equation
\PF\ admits a constant solution $f^x_0=1$ which is defined everywhere.
Before presenting the details of the other solutions, we would like to
discuss some of their general properties.

The solutions to the indicial equations at the three singular points are
$(0,0,0)$ at $z=0$, $(0,1,1)$ at $z=1$ and
$\left(0,{1\over 3},{2\over 3}\right)$ at $z=\infty$.
Accordingly, we expect the solutions near $z=0$ to be logarithmic.
At $z=1$ we expect a power series solution with index $1$ and a
logarithmic solution, while at $z=\infty$ both solutions are expected
to be power series involving fractional powers. Therefore we will
obtain nontrivial monodromy transformations about these points.
Moreover, the fact that all solutions at $z=\infty$ are power series
signals an exactly solvable conformal field theory associated to that
point. This is the perturbative orbifold point, similar to the
Gepner points in the one
parameter models considered in
\nref\CDGP{P. Candelas, X.C. De La Osa, P.S. Green and L. Parkes,
``A Pair of Calabi-Yau Manifolds as An Exactly Soluble Superconformal
Theory'', Nucl. Phys. {\bf B359} (1991) 21.}%
\nref\KT{A. Klemm and S. Theisen, ``Considerations of One Modulus
Calabi-Yau Compactifications: Picard-Fuchs Equations, \ka\ Potentials
and Mirror Maps'', \np{389}{1993}{153}, hep-th/9205041.}%
\nref\AF{A. Font, ``Periods and Duality Symmetries in Calabi-Yau
Compactifiactions'', \np{391}{1993}{358}, hep-th/9203084.}%
\refs{\CDGP,\KT,\AF}.

In the context of global mirror symmetry, it is known that one can define
a special basis of solutions such that the monodromy transformations
are integral symplectic matrices. Using special coordinates $t^a$ on the
moduli space, this basis can be written in general as
$\left(1,t^a,{\del F\over \del t^a},2F-t^a{\del F\over \del t^a}\right)$
where $F(t^a)$ is the $N=2$ prepotential. The functions
$(t^a,{\del F\over \del t^a})$ represent the periods of the holomorphic
three-form of the mirror manifold with respect to a symplectic basis
of cycles.

In the local case, we have only three periods, therefore
we cannot find a symplectic basis. This is consistent with the
local mirror geometry \ellfiber\ being one dimensional. In fact, as
observed in \refs{\AGM,\KZ}, the equation \PF\ is the
logarithmic integral of an ordinary hypergeometric equation
\eqn\ordhyper{
\left[\theta_z^2-z\left(\theta_z+{1\over 3}\right)
\left(\theta_z+{2\over 3}\right)\right]f=0.}
This represents the Picard-Fuchs equation for the
periods of the global holomorphic one-form on a symplectic
basis of cycles of the elliptic curve \ellfiber. As before, we
have regular
singular points at $(0,1,\infty)$,
the monodromy group being isomorphic to
$\Gamma (3)$ \ref\IKSY{K. Iwasaki, H. Kimura,
S. Shimomura and M. Yoshida, ``From Gauss to Painlev\'e --
A Modern Theory of
Special Functions'', Aspects of Mathematics, Vieweg, 1991.}.
Therefore we can define a
 special basis of solutions of the form $(1,t(z),t_d(z))$ \KZ\
such that the monodromy transformations are integral $3\times 3$
matrices of the form
\eqn\monform{
\left[\matrix{&1&0\cr &m&M\cr}\right].}
Here $M\in \Gamma (3)$ represents the $\Gamma(3)$ monodromy acting
on $(t,t_d)$ which are logarithmic integrals of the solutions of
\ordhyper. The integral column vector $m$ reflects the fact that
the periods $(t,t_d)$ may pick up constant shifts when analytically
continued about the singular points. This will play an important role
at a latter stage.
We now derive explicit formulae for the local solutions and determine
the monodromy.

$i)Solutions\ at\ z=0$.

The other two solutions
in the region $|z|<1$ have been given in integral form in \Mj.
We have one logarithmic solution $f^0_1(z)=G^{2,2}_{3,3}
\left(-z\left|\matrix{& {1\over 3}&{2\over 3}&{1}\cr &0 &0 &0\cr}
\right.\right)$ which can be represented in integral form\foot{This
function differs by a normalization factor $-{1\over
\gm{1\over 3}\gm{2\over 3}}$ from the original G-function of \Mj.
We have included this factor for latter convenience.}
\eqn\logsol{
G^{2,2}_{3,3}
\left(-z\left|\matrix{& {1\over 3}&{2\over 3}&{1}\cr &0 &0 &0\cr}
\right.\right)={1\over 2\pi i\gm{1\over 3}\gm{2\over 3}}
\int_C{\gm{-s}\gm{s+{1\over 3}}\gm{s+{2\over 3}}\over s\gm{1+s}}(-z)^sds}
where $|z|<1$ and $|\hbox{arg}(z)|<\pi$.
The contour $C$ is parallel with the imaginary axis, encircles the
origin and closes to the right as shown in the figure below.

\ifig\cont{The integration contour for Meijer's G-functions.}
{\epsfxsize2.5in\epsfbox{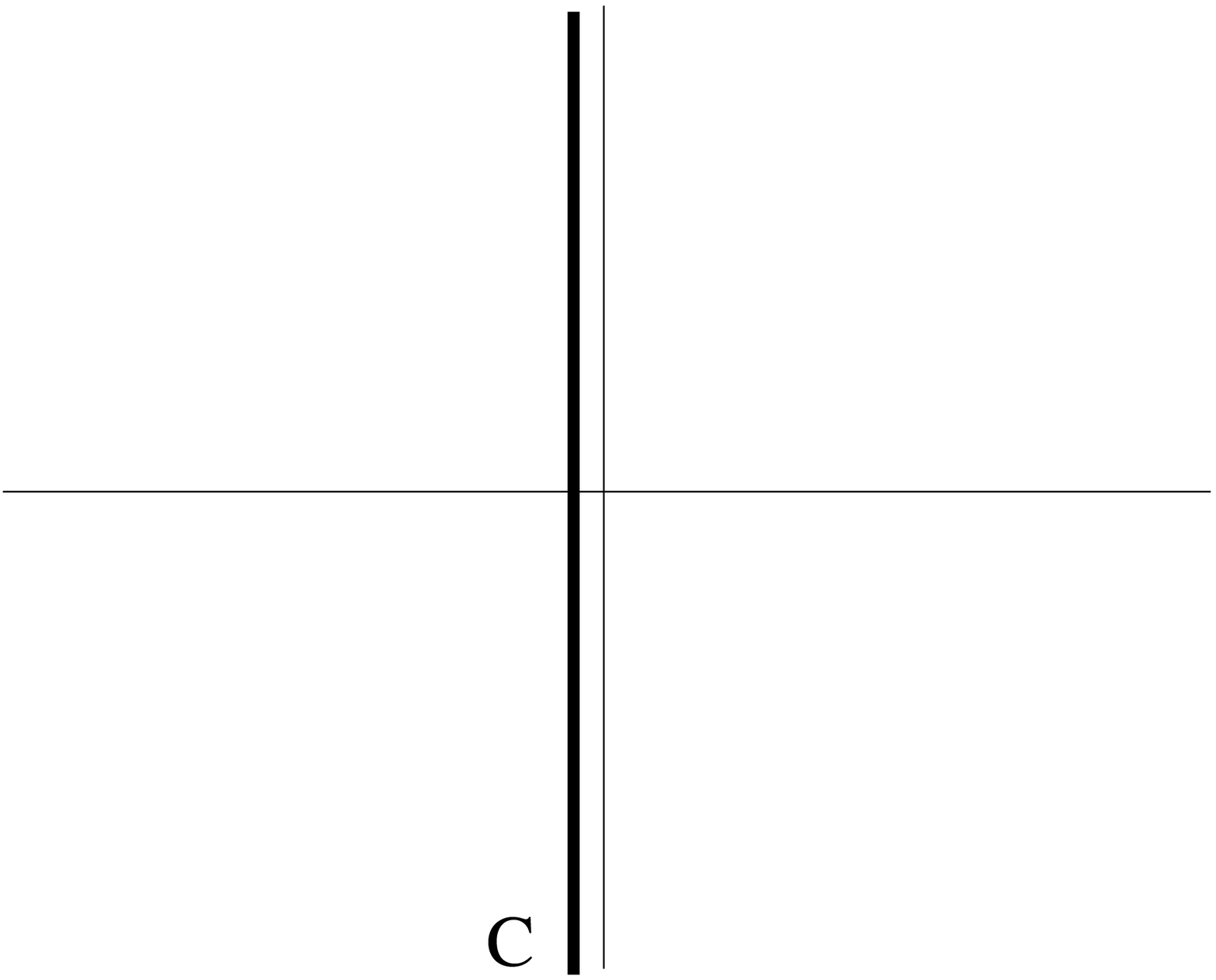}}

\noindent
By evaluating the residues, we can rewrite \logsol\ as a power series
\eqn\seriesA{
f^0_1(z)=
\lg\left(-{z\over 27}\right)+{1\over  \gm{1\over 3}\gm{2\over 3}}
\sum_{n=1}^{\infty}{\gm{n+{1\over 3}}\gm{n+{2\over 3}}\over n\gm{1+n}^2}
z^n.}
On general grounds, we expect one extra independent solution in order
to obtain a complete system. It turns out that there exist
two particular double logarithmic solutions\foot{These solutions differ
by a $-$ sign from the original G-functions of \Mj.}
\eqn\doublog{\eqalign{
& G_{3,3}^{3,1}\left(-z\left|\matrix{& {1\over 3}&{2\over 3}&{1}\cr
&0 &0 &0\cr}\right.\right)
={1\over 2\pi i}\int_C
{\gm{-s}^2\gm{{2\over 3}+s}\over s\gm{{2\over 3}-s}}(-z)^sds\cr
& G_{3,3}^{3,1}\left(-z\left|\matrix{& {2\over 3}&{1\over 3}&{1}\cr
&0 &0 &0\cr}\right.\right)
={1\over 2\pi i}\int_C
{\gm{-s}^2\gm{{1\over 3}+s}\over s\gm{{1\over 3}-s}}(-z)^sds\cr}}
where the integration is taken along the same contour $C$ as above
and $|z|<1$ and $|\hbox{arg}(z)|<\pi$.
Adopting the notation of \Mj,\ these functions will be denoted by
$G_{3,3}^{3,1}\left(-z\left|\left|{1\over 3}\right.\right.\right)$ and
respectively
$G_{3,3}^{3,1}\left(-z\left|\left|{2\over 3}\right.\right.\right)$.
Again, we can evaluate the residues and obtain the following series
expansions
\eqn\seriesB{\eqalign{
& G_{3,3}^{3,1}\left(-z\left|\left|{1\over 3}\right.\right.\right)=
{1\over 2}
\lg^2\left(-{z\over 27}\right)+{\pi\sqrt{3}\over 3}
\lg\left(-{z\over 27}\right)+{\pi^2\over 3}+\cr
& {1\over \gm{1\over 3}
\gm{2\over 3}}\sum_{n=1}^{\infty}{\gm{n+{1\over 3}}\gm{n+{2\over 3}}
\over n\gm{1+n}^2}z^n\lg\left(-{z\over 27}\right)+\cr
& {1\over \gm{1\over 3}\gm{2\over 3}}\sum_{n=1}^{\infty}
\left[n\left(\dig{{2\over 3}+n}+\dig{{2\over 3}-n}-2\dig{1+n}+3\lg{3}
\right)-1\right]\times \cr
&{\gm{{1\over 3}+n}\gm{{2\over 3}+n}\over n^2\gm{1+n}^2}
z^n.\cr}}
\eqn\seriesC{\eqalign{
&G_{3,3}^{3,1}\left(-z\left|\left|{2\over 3}\right.\right.\right)=
{1\over 2}
\lg^2\left(-{z\over 27}\right)-{\pi\sqrt{3}\over 3}
\lg\left(-{z\over 27}\right)+{\pi^2\over 3}+\cr
& {1\over \gm{1\over 3}
\gm{2\over 3}}\sum_{n=1}^{\infty}{\gm{n+{1\over 3}}\gm{n+{2\over 3}}
\over n\gm{1+n}^2}z^n\lg\left(-{z\over 27}\right)+\cr
& {1\over \gm{1\over 3}\gm{2\over 3}}\sum_{n=1}^{\infty}
\left[n\left(\dig{{1\over 3}+n}+\dig{{1\over 3}-n}-2\dig{1+n}+3\lg{3}
\right)-1\right]\times \cr
&{\gm{{1\over 3}+n}\gm{{2\over 3}+n}\over n^2\gm{1+n}^2}
z^n.\cr}}
Note that the three solutions presented in \seriesA\ and \seriesB\
are not independent. They satisfy the linear dependence relation
\eqn\lindep{
\gm{1\over 3}\gm{2\over 3}G^{2,2}_{3,3}
\left(-z\left|\matrix{& {1\over 3}&{2\over 3}&{1}\cr &0 &0 &0\cr}
\right.\right)=
G_{3,3}^{3,1}\left(-z\left|\left|{1\over 3}\right.\right.\right)-
G_{3,3}^{3,1}\left(-z\left|\left|{2\over 3}\right.\right.\right).}
A general local solution near $z=0$ can therefore be written as
any linear combination of these three functions subject to the
constraint \lindep.
For latter convenience, we will consider the following linear
combination
\eqn\lincomb{
f^0_2(z)={1\over 2}\left(G_{3,3}^{3,1}\left(-z\left|\left|{1\over 3}
\right.\right.\right)+G_{3,3}^{3,1}\left(-z\left|\left|{2\over 3}\right.
\right.\right)\right).}
This can be rewritten in the form
\eqn\tform{
f^0_2(z)={1\over 2}f^0_1(z)^2-{1\over 12}+O(z).}

Note that the above solutions are
absolutely convergent in the region $|z|<1$. If $|z|=1$ the series
are convergent for $z\neq 1$. However, the simple logarithmic solution
\logsol\ diverges at $z=1$,
which is a singular point  similar to the conifold point in the
quintic moduli space \CDGP.
This singular point  will play an important role in
determining the integral basis of solutions $(1,t,t_d)$.
According to \KZ,\ the
coordinate $t$ can be identified with the simple logarithmic solution
\seriesA.\ More precisely, we will take
\eqn\speccoord{
t(z)={1\over 2\pi i}f^0_1(z).}
The dual period $t_d={\del F\over \del t}$ can be found by
studying the system of solutions near the singular point $z=1$ and
analytically continuing the vanishing period \CDGP.

$ii)Solutions\ at\ z=1$.

A basis of solutions near this singular point can be found by changing
variables $u=1-z$ in \PF.\ This results in the following differential
equation
\eqn\conifeq{
\left(\theta_u^3-{2+u\over 1-u}\theta_u^2+{1\over 9}
{2u^2-2u+9\over (1-u)^2}\theta_u\right)f=0}
which can be solved with standard
recursive methods (see for example \IKSY.)\
As mentioned before, the solutions of the indicial equation are
$(0,1,1)$, therefore one of the solutions should be a power
series of index one, while the second solution should be logarithmic.
It can be shown \KZ\ that the logarithmic solution is given by the
analytic continuation of the period $t(z)$ to $z=1$
\eqn\seriesD{
t(u)=-3t_d(u)\lg(u)+O(1).}
The function
\eqn\vanper{
t_d(u)=-{\sqrt{3}\over 6}\left(u+{11\over 18}u^2+\ldots\right)}
is itself a solution of the equation and
represents the vanishing period at the singular
point. It is related by analytic continuation to the dual period
$t_d(z)$. Using these facts, one can find the precise form of the
period $t_d(z)$ near $z=0$
\eqn\dual{\eqalign{
t_d(z)&=-{1\over 4\pi^2}f^0_2(z)-{1\over 2}t(z)+{1\over 3}\cr
&={1\over 2}t(z)^2-{1\over 2}t(z)+{1\over 4}+O(z)}}
where $f^0_2(z)$ has been defined in \lincomb.\

$iii)Solutions\ at\ z=\infty$.

The local solutions in the region $|z|>1$ can be found
similarly, by changing variables $\zeta=1/z$ and then solving
the resulting equation
\eqn\orbeq{
\left[\zeta{\theta_\zeta}^3-\left(\theta_\zeta-{1\over 3}\right)
\left(\theta_\zeta-{2\over 3}\right)\theta_\zeta\right]f=0.}
This is again a Meijer equation whose local solutions can be determined
by analytically continuing the functions
$G_{3,3}^{3,1}\left(-z\left|\left|{1\over 3}
\right.\right.\right)$ and $G_{3,3}^{3,1}
\left(-z\left|\left|{2\over 3}\right.
\right.\right)$ in a neighborhood of $z=\infty$ \Mj. We obtain
$f^\infty_1(\zeta)=E_{3,3}\left(-{1\over\zeta}\left|\left|{2\over 3}
\right.\right.\right)$, $f^\infty_2(\zeta)=E_{3,3}
\left(-{1\over\zeta}\left|\left|{1\over 3}\right.\right.\right)$, where
\eqn\locsol{\eqalign{
&E_{3,3}\left(-{1\over\zeta}\left|\left|{2\over 3}
\right.\right.\right)=
-3{\gm{1\over 3}^2\over\gm{2\over 3}}e^{\pi i\over 3}{\zeta}^{{1\over 3}}
{}_3F_2\left({1\over 3},{1\over 3},{1\over 3};
{2\over 3},{4\over 3};{\zeta}\right)\cr
&E_{3,3}\left(-{1\over \zeta}\left|\left|{1\over 3}\right.\right.\right)
=-{9\over 2}{\gm{2\over 3}^2\over \gm{1\over 3}}
e^{2\pi i\over 3}{\zeta}^{{2\over 3}}
{}_3F_2\left({2\over 3},{2\over 3},{2\over 3};
{4\over 3},{5\over 3};{\zeta}\right).\cr}}
These solutions are well defined and convergent if $|\zeta|<1$ and
$|\hbox{arg}(\zeta)|<\pi$.

It follows that the analytic continuation of the periods $(t,t_d)$
in a neighborhood of $\infty$ reads
\eqn\orbbasis{\eqalign{
&t(\zeta)={{\omega}^2-\omega\over 4{\pi}^2}
\left(E_{3,3}\left(-{1\over \zeta}\left|\left|{2\over 3}
\right.\right.\right)-
E_{3,3}\left(-{1\over \zeta}\left|\left|{1\over 3}\right.
\right.\right)\right)\cr
&t_d(\zeta)={1\over 3}+{1\over 4{\pi}^2}
\left(\omega E_{3,3}\left(-{1\over \zeta}\left|\left|{2\over 3}
\right.\right.\right)+{\omega}^2
E_{3,3}\left(-{1\over \zeta}
\left|\left|{1\over 3}\right.\right.\right)\right)\cr}}
where $\omega=e^{2\pi i\over 3}$.
This can be proved by direct computation using
\speccoord,\ \dual,\ and \lindep\ and taking into account
the fact that ${1\over {2\pi i \gm{1\over 3}\gm{2\over 3}}}=
{{\omega}^2-\omega\over 4{\pi}^2}$.
Note that the values of the periods at the singular point are
\eqn\values{
t\left(\zeta=0\right)=0,\qquad t_d\left(\zeta=0\right)={1\over 3}.}

Once we know the local solutions near each singular point, the global
solution can be obtained by patching them together. Since the periods
are multivalued functions, this process requires branch cuts in the
complex plane. These are implicit in the restrictions imposed
on the phase of $z$ in the paragraphs following \logsol\ and \locsol\
which represent two branch cuts joining $z=0$ and $z=1$ and
respectively $z=1$ and $z=\infty$ along the real axis. This will allow
us to assign unambiguous values of the periods to all points in the
moduli space (away from the cuts).

{\it iv)Monodromy}.

The monodromy of the integral basis of solutions around the
singular points can be determined from the explicit local solutions
derived above. By performing a counterclockwise rotation about each
of the three singular points, we find
\eqn\monodromy{
M_0=\left[\matrix{&1 &0 &0\cr &1 &1 &0\cr &0 &1 &1\cr}\right]
\qquad
M_1=\left[\matrix{&1 &0 &0\cr &0 &1 &-3\cr &0 &0 &1\cr}\right]
\qquad
M_\infty=\left[\matrix{&1 &0 &0\cr &1 &-2 &-3\cr &0 & 1 &1\cr}\right]}
satisfying $M_\infty=M_1M_0$.
Note that $M_0$ and $M_1$ are of infinite order and they agree with
the expected behavior of the periods in the infinite volume limit and
respectively at the conifold point. However, $M_\infty$ is of third order,
reflecting the quantum ${{\bf Z}_3}$ symmetry of the orbifold point as explained
in \A. This identifies the orbifold point as a quotient singularity
in the moduli space.

\subsec{Fractional Branes and BPS States on The Moduli Space}

We now show how to use the above detailed solution in order to
relate the orbifold fractional branes to the generic BPS states
on the moduli space. This will allow us to deform this states
along a path between the orbifold point and the large radius limit
and to finally interpret them as macroscopic branes on supersymmetric
cycles.

In order to carry out this program, we need to know the form of the
complexified \ka\ class $t_b\equiv B+iJ$ in terms of the exact periods
$(1,t,t_d)$. This is essentially the problem of finding a precise form
of the mirror map. According to the properties of the monomial-divisor
map as discussed for example in \AGM,\ the asymptotic form of
$t_b$ is
\eqn\asBfield{
t_b\sim {1\over 2\pi i}\lg{\left(\pm{z\over 27}\right)+O(z)}.}
Since the period $t$ has an asymptotic behavior
$\sim \lg{\left(-{z\over 27}\right)}$ this fixes $t_b=t+C$,
where $C$ is an integration constant which depends on the choice
of sign of $z/27$ inside the logarithm. It has been argued in
\AGM\ that this sign should be positive and $C$ was fixed to
$-{1\over 2}$
\eqn\Bfield{
t_b=t-{1\over 2}.}
With this choice for the mirror map, the quantum volume of the
${\bf P}^1$ cycle in the exceptional divisor is positive everywhere
on the moduli space. In particular the value of $t_b$ at the orbifold
point is $1/2$.\foot{Note that fixing the integration constant
is physically meaningful. We will show later that the present choice 
corresponds to turning on a half integral 
B field on the hyperplane cycle of ${\bf P}^2$.}.

Note that, in the basis $(1,t_b,t_d)$, the monodromy matrices
will no longer be integral since $t_b$ and $t_d$ are shifted by
half integral numbers.
Therefore, the BPS states will be represented with respect to the
basis $(1,t_b,t_d)$ by charge vectors $(n_0,n_1,n_2)$ where
$n_1,n_2$ are integral and $n_0$ is half integral.
The associated central charge is
\eqn\central{
Z(n_0,n_1,n_2)=n_0+n_1t_b+n_2t_d.}
Our first goal is to identify the charge vectors associated
with the three boundary states constructed in the previous section.
Then, using analytic continuation to the large radius limit, we will
interpret the resulting BPS states as D-branes wrapped on cycles.

The first step is to note that the perturbative ${{\bf C}^3}/{{\bf Z}_3}$
orbifold CFT has a ${{\bf Z}_3}$ quantum symmetry \A\ which permutes the
twist fields cyclically. Taking into account the boundary state
construction in the previous sections, the quantum ${{\bf Z}_3}$ also
permutes the fractional branes in a similar manner. Therefore
all boundary states, associated with the three irreducible
representations of ${\bf Z}_3$ $\gamma_1,\gamma_2$ and $\gamma_3$,
form an orbit of the discrete global symmetry group.
Changing perspective, the same quantum symmetry manifests itself
in the form of a third order monodromy of periods on the moduli space
as explained in \A\ and in the previous paragraph. From here we
conclude that the fractional branes are naturally in
one to one correspondence with a set of three periods forming an orbit
of the monodromy generator $M_\infty$.

The next available piece of information is that all fractional branes
have equal mass, which is $1/3$ of the mass of a $D0$-brane. At the same
time the formula \values\ shows that this is precisely the mass of a
state with charges $(0,0,\pm 1)$ at the orbifold. Therefore we will
identify the three fractional branes with the following states\foot{The
above arguments do not fix the sign of the charges. We have chosen the
sign so that the three states will have a total D0-brane charge $1$
rather than $-1$, as it will be clear latter. A different choice of sign
would correspond to antiparticle states.}
\eqn\states{
(0,0,1),\qquad \left({1\over 2},1,1\right),\qquad
\left({1\over 2},-1,-2\right)}
obtained by acting\foot{Note that the monodromy matrices act on the
charge vectors by right multiplication and on the period vectors by left
multiplication.} by the monodromy generator $M_{\infty}$.
The corresponding central charges read
\eqn\ctrcharges{
Z(0,0,1)=t_d,\qquad Z\left({1\over 2},1,1\right) ={1\over 2}+t_b+t_d,
\qquad
Z\left({1\over 2},-1,-2\right)={1\over 2}-t_b-2t_d.}
Clearly, all states have mass equal to $1/3$ of the D0-brane mass
which is normalized to one.

In order to complete the analysis, we have to understand if the proposed
BPS states have a well defined D-brane interpretation in the large
radius limit. Although the periods can be continued along any
given path joining the two points, the continuation of
the BPS states is more subtle due to the possible jumping phenomena.
This phenomenon and the associated marginal stability curves have been
studied intensively in the context of Seiberg-Witten solutions
\nref\SW{N. Seiberg and E. Witten, ``Electric-Magnetic Duality, Monopole
Condensation and Confinement in $N=2$ Supersymmetric Yang-Mills theory'',
Nucl. Phys. {\bf B426} (1994) 19, hep-th/9407087.}%
\nref\FB{F. Ferrari and A. Bilal, ``The Strong Coupling Spectrum of The
Seiberg-Witten Theory'', \np{469}{1996}{387}, hep-th/9602082.}%
\nref\WL{W. Lerche, ``Introduction to Seiberg-Witten Theory and Its
Stringy Origin'', Nucl. Phys. {\bf B} (Proc. Suppl.) {\bf 55} (1997) 83,
hep-th/9611190.}%
\refs{\SW,\FB,\WL}.
In the present case, note that all periods have real values at the
orbifold point, therefore all marginal stability curves will necessarily
pass through that point. Therefore, as long as the curves are reasonably
shaped (that is if they have no self intersection), it is possible to
find a path between the
orbifold point and the large radius limit that avoids them. In the
following we will assume that this is in fact the case and that there
exists such a path along which the states \states\ are stable.
The final result will be shown to be consistent with this assumption.

The asymptotic expansion of the central
charges of the three BPS states reads
\eqn\asympch{\eqalign{
&Z(0,0,1)={1\over 2}t_b^2+{1\over 8}+O(z)\cr
&Z({1\over 2},1,1)={1\over 2}t_b^2+t_b+{5\over 8}+O(z)\cr
&Z({1\over 2},-1,-2)=-t_b^2-t_b+{1\over 4}+O(z).\cr}}

The corresponding D-brane states can be identified by comparing
\asympch\ to the semiclassical expression for
the central charge of a state with effective D-brane charges
$(q_0,q_2,q_4)$
\eqn\Dcentral{
Z(q_0,q_2,q_4)=-{q_4}{t_b^2\over 2}+q_2t_b+q_0.}
Note that, according to the discussion in section four, the D-brane
states are classified by the K-theory group $K(D)$, where
$D\simeq {\bf P}^2$ is the exceptional divisor.
The vector $Q=(q_0,q_2,q_4)$ takes values in the
total cohomology space $H^4\left(D,{\bf Q}\right)\oplus
H^2\left(D,{\bf Q}\right)\oplus H^0\left(D,{\bf Q}\right)$
and it measures the effective charges of a brane configuration
represented by a given K-theory class. We consider rational
cohomology since the effective charges may be fractional.
Given a K-theory class represented by a vector bundle
(or, more generally, a coherent sheaf) $V$ on
$D$, we can determine $Q$ from the Chern-Simons couplings found in
\nref\D{M.R. Douglas, ``Branes within Branes'', contributed to
``Cargese 1997, Strings, Branes and Dualities'', 267-275,
hep-th/9512077.}%
\nref\GHM{M. Green, J.A. Harvey and G. Moore,
``I-Brane Inflow and Anomalous Couplings on D-Branes'',
Class. Quant. Grav. {\bf 14} (1997) 47, hep-th/9605033.}%
\nref\HM{J.A. Harvey and G. Moore, ``On The Algebras of BPS States'',
Commun. Math. Phys. {\bf 197} (1998) 489, hep-th/9609017.}%
\refs{\D,\GHM,\HM}.
A careful analysis of these couplings, taking into account the
twisting of the worldvolume fermions in normal directions,
has been performed in 
\ref\CY{Y.-K E.Cheung and Z. Yin, ``Anomalies, Branes and Currents'',
Nucl. Phys. {\bf B517} (1998) 69, hep-th/9710206.}.
\refs{\MM,\CY}
According to the results therein, the vector of induced charges for a 
system of D$p$-branes
wrapping a supersymmetric cycle $D$ is given by
\eqn\induced{
Q=\hbox{ch}(V)\sqrt{{\hat A}(T_D)\over {\hat A}(N_D)}.}
Here $T_D$ and $N_D$ denote the tangent and respectively the 
normal bundle
to the cycle $D$\foot{There is a subtlety 
related to this formula which has been clarified in 
\ref\FW{D. Freed and E. Witten, 
``Anomalies in String Theory with D-Branes'', 
hep-th/9907189.}. Since ${\bf P}^2$ is not a spin manifold, but only a 
$\hbox{spin}^c$-manifold, the bundle $V$ on $D$ is actually a 
$\hbox{spin}^c$-bundle. This means that the curvature $\hbox{Tr}(F)$ is a 
half-integral class rather than integral, as in the conventional case. 
However, in the present case, we claim that there is a flat background 
B-field turned on such that $\int_{{\bf P}^1}B\in {\bf Z}+{1\over 2}$.
The presence of this B-field is related to the choice of the integration 
constant of the mirror map discussed after equation (6.33). 
The net effect is 
to cancel the effect of $w_2(D)$ 
discussed in \FW,\ so that $V$ can be regarded as a 
conventional vector bundle in \induced.}

For the case when $D$ is a holomorphic surface embedded in Calabi-Yau
threefold, we have \CY\
\eqn\Agenus{\eqalign{
\sqrt{{\hat A}(T_D)\over {\hat A}(N_D)}&=1+{1\over 48}\left(p_1(N_D)-
p_1(T_D)\right)\cr
&=1+{\chi(D)\over 24}w_D\cr}}
where $\chi(D)$ is the topological Euler characteristic of $D$
and $w_D$ is the fundamental class. Therefore we obtain
\eqn\effch{
Q=r+c_1(V)+\left({r\over 8}+{1\over 2}c_1^2(V)-c_2(V)\right)w_D.}
Using this formula, we can show that the D-brane configurations
corresponding to the three central charges \asympch\ are
\eqn\corresp{\eqalign{
& Z(0,0,1)\rightarrow {\overline {D4}}\cr
& Z({1\over 2},1,1)\rightarrow {\overline {D4}}+D2\cr
& Z({1\over 2},-1,-2)\rightarrow 2D4+{\overline {D2}}+D0.\cr}}
In the above, the symbol $D4$ represents a D4-brane wrapped on the
exceptional divisor $D$ while $D2$ represents a D2-brane wrapped on
a ${\bf P}^1\subset {\bf P}^2$ cycle in the hyperplane class $H$.
Barred symbols denote antibrane states which correspond to ``negative''
K-theory classes. These states correspond to the fractional branes
at the orbifold.

The first two configurations
are represented by the classes $-[\co(-1)]$ and $-[\co]$ where
$\co$, $\co(-1)$ are the trivial and respectively the tautological
line bundle on ${\bf P}^2$. The third case deserves more
attention since it corresponds to a rank 2 holomorphic bundle
$V$ on ${\bf P}^2$ characterized by
\eqn\bundle{
r=2,\qquad c_1(V)=-1,\qquad c_2(V)=1.}
As a consistency check we should now be able to check if the
resulting D-brane states are indeed BPS. According to  \HM,\
this means that the three bundles should be holomorphic and stable.
These criteria are clearly satisfied by the first two line bundles.
However, this is a nontrivial test for the rank 2 bundle $V$
\nref\DL{J.-M. Dr\'ezet and J. Le Poitier, ``Fibr\'es Stable et
fibr\'es Exceptionnels sur Le Plan Projectif'', Ann. Scient. Ec Norm. Sup.
{\bf 18} (1998) 105.}%
\nref\L{J. Le Poitier, ``Lectures on Vector Bundles'', Cambridge
University Press, 1997.}%
\refs{\DL,\L}.
According to the classification therein, it can be checked that the
bundle $V$ in our problem is an exceptional stable holomorphic bundle
on ${\bf P}^2$. This means that it is holomorphic and stable and it has
no deformations i.e. the moduli space reduces to a single point.
It is remarkable that our BPS states analysis has lead precisely
to one of these exceptional bundles which form a special discrete
series. Moreover, note that both the fractional branes and the
D-brane configurations \corresp\ have no moduli. This is in agreement
with the arguments of \BD\ which show that the number of moduli of
the BPS states should be preserved under \ka\ deformations.

To this end, note that there is one more test we can perform.
In \DF,\ Douglas and Fiol have introduced the index
\eqn\index{
\hbox{tr}_{{\cal H}_{open}}\left((-1)^Fe^{-2tH_o}\right)}
that counts the number of fermion zero modes in the Ramond open string
sector in the presence of D-brane states. They have also argued on the
basis of Dirac quantization condition, that this index should actually
compute the classical intersection number of supersymmetric cycles up
to sign. More precisely, the index in \index\ can be given a closed
string interpretation using the formalism of boundary states discussed
previously. In fact \index\ is related by a modular transformation
to the following closed string amplitude
\eqn\closedind{
{}_{RR}\langle B_1|e^{-lH_{c}}|B_2\rangle_{RR}}
where $|B_{1,2}\rangle$ are two arbitrary boundary states.
This formula defines an antisymmetric bilinear form on the set of
all boundary states which is the equivalent of an intersection form.
In the case of even branes, this seems to lead to a puzzle since
the classical intersection form is symmetric. The resolution resides
in the fact that the open string index computes the intersection form
of the branes as seen from the dual mirror point of view \BD. The
exact periods $(t_b,t_d)$ found in the previous subsection, are really
sections of a rank two holomorphic vector bundle $E$ on the moduli space
with structure group $\Gamma(3)\subset SL(2,{\bf Z})$. The charge
$(n_1,n_2)$ vectors of BPS states are locally constant sections
and we can define a natural symplectic form
\eqn\symplectic{\omega :((n_1,n_2),(n_1^\prime,n_2^\prime))\rightarrow
3(n_1^\prime n_2-n_1n_2^\prime ).}
This represents the intersection form
on the middle homology of the elliptic curve \ellfiber. The coefficient
$3$ has been chosen to agree with the intersection number $H\cdot D=-3$.

In our situation the index \index\ can be easily
computed at the orbifold point using conformal field theory techniques.
As noted in \DF,\ it turns out that for a given pair of fractional branes
one actually counts the net number of chiral fermion multiplets in
the RR sector of the open string stretching between the branes which
are left invariant by the orbifold projection.
Therefore, in the presence of two fractional branes classified by two
irreducible representations $\gamma_a$, $\gamma_b$, we count the
number of chiral
fermion multiplets $\chi$ satisfying
\eqn\invar{
\gamma(g)^{-1}_a\chi \gamma(g)_b=R(g)\chi}
where $g$ is a generator of ${{\bf Z}_3}$ and $R$ is the standard three
dimensional representation defined by embedding in $SU(3)$.
The result has a concise
graphical description encoded in the quiver diagram
corresponding to the regular representation.

\ifig\cont{The quiver diagram for the regular representation of
${\bf C}^3/{\bf Z}_3$.}
{\epsfxsize1.7in\epsfbox{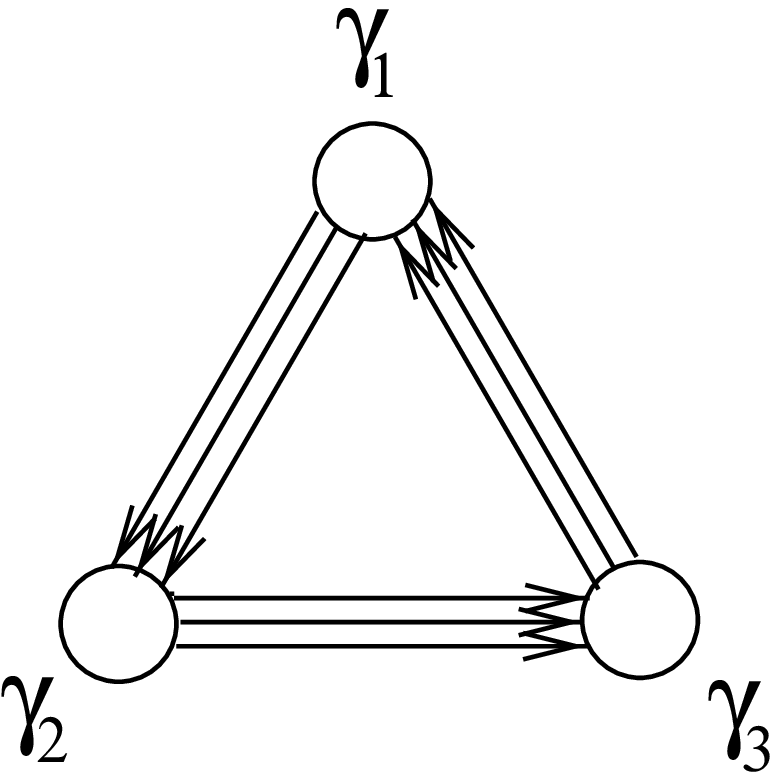}}

Given two irreducible
representations $\gamma_a$, $\gamma_b$ the $(a,b)$-th
entry of the resulting
intersection form is given by the number of edges connecting the two
vertices. The sign is given by the orientation of the edges:  the
contribution is positive if the arrows point
from the vertex $a$ to the vertex $b$ and negative in the reversed
situation. This gives the following antisymmetric intersection form

 $$\vbox{\offinterlineskip
 \halign{&\hfil#\hfil\vrule\quad&\strut\hfil#\hfil\quad&\strut\hfil#\hfil\quad
 &\strut\hfil#\hfil \cr & $\gamma_1$ & $\gamma_2$ & $\gamma_3$ \cr 
\noalign{\hrule} \cr
  $\gamma_1$ & $0$ & $3$ & \hskip-7pt$-3$ \cr
         $\gamma_2$ & \hskip-7pt$-3$ & $0$ & $3$ \cr
         $\gamma_3$ & $3$ & \hskip-7pt$-3$ & $0$ \cr}}$$%
An elementary computation shows that this agrees with the intersection
form \symplectic\ evaluated on the charges \states.

\vfill\eject

\centerline{\bf Acknowledgments}

It is a pleasure to thank Micha Berkooz, Ilka Brunner, Paul Cohen, 
Schuyler Cullen,
Rami Entin, Bartomeu Fiol,
Albrecht Klemm, Greg Moore, Christian R\"omelsberger, Moshe Rozali,
Steve Shenker, Eva Silverstein, Yun Song,  John Tate, and especially
Michael Douglas
for very useful discussions
and suggestions. We would like to thank Rami Entin for collaboration at
an early stage of the project and Albrecht Klemm for collaboration
in the results of section 6.1. The work of D.-E. D. has been supported by
DOE grant DE-FG02-90ER40542. The work of J.G. has been supported
by Rutgers University and Stanford University with NSF grant
PHY-9870115.

\vfill\eject
\appendix{A}{Conformal Field Theory Conventions}

In this appendix we summarize some of the conventions we have used
as well as properties of $\vartheta$ functions that are needed for
the construction of the boundary states. For the flat space
discussion we have used the following mode expansions for the open
string fields
\eqn\modeex{\eqalign{
X^\mu&=x^\mu+2\pi p^\mu t+i\sqrt{2}\sum_{n\neq 0}{1\over
n}\alpha_n^\mu e^{-i\pi nt}\cos(n\pi \sigma)\qquad
\mu=0,\ldots,p\cr
\psi^M&=\sqrt{\pi}\sum_r\psi_r^Me^{-i\pi r(t-\sigma)}\qquad
\qquad \qquad \qquad \qquad \quad M=0,\ldots,9\cr
\wt\psi^\mu&=\sqrt{\pi}\sum_r\psi_r^\mu e^{-i\pi r(t+\sigma)}\cr
X^i&=i\sqrt{2}\sum_{n\neq 0}{1\over
n}\alpha_n^ie^{-i\pi nt}\sin(n\pi \sigma)\qquad
\qquad \qquad \qquad \quad i=p+1,\ldots,9\cr
\wt\psi^i&=-\sqrt{\pi}\sum_r\psi_r^i e^{-i\pi r(t+\sigma)}\cr},}
where as usual $r\in Z$ in the R sector and $r\in Z+1/2$ in the NS
sector and $0\leq\sigma\leq 1$. For the closed string the expansions are
\eqn\closedexp{\eqalign{
X^M&=x^M+2\pi p^\mu t+{i \over \sqrt{2}}\sum_{n\neq 0}({1\over
n}\alpha_n^M e^{-2\pi i n(t-\sigma)}+\wt\alpha_n^M e^{-2\pi i
n(t+\sigma)})\qquad M=0,\ldots,9\cr
\psi^M&=\sqrt{2\pi}\sum_r\psi_r^Me^{-2\pi i r(t-\sigma)}\cr
\wt\psi^M&=\sqrt{2\pi}\sum_r\psi_r^M e^{-2\pi i r(t+\sigma)}},}
and the oscillator algebra is as usual
\eqn\coomm{\eqalign{
[\alpha_n^M,\alpha_m^N]&=[\wt\alpha_n^M,\wt\alpha_m^N]
=n\delta_{n+m}\delta_{MN}\cr
\{\psi^M_r,\psi^N_s\}&=\{\wt\psi^M_r,\wt\psi^N_s\}
=\delta_{r+s}\delta_{MN}}.}

The action of the orbifold group $\Gamma$ is natural on the
complex coordinates \eqn\comp{\eqalign{ Z^i&={1\over
\sqrt{2}}(X^{2i}+ iX^{2i+1})\qquad i=1,2,3\cr \bar{Z}^i&={1\over
\sqrt{2}}(X^{2i}- iX^{2i+1})\cr \lambda^i&={1\over
\sqrt{2}}(\psi^{2i}+i\psi^{2i+1})\cr \bar{\lambda}^i&={1\over
\sqrt{2}}(\psi^{2i}- i\psi^{2i+1})},} so that the oscillator
expansion of these fields in the the closed string is
\eqn\closedexpb{\eqalign{ Z^i&=z^i+2\pi p^i t+{i \over
\sqrt{2}}\sum_{n\neq 0}({1\over n}\beta_n^i e^{-2\pi i
n(t-\sigma)}+\wt\beta_n^i e^{-2\pi i n(t+\sigma)})\cr
\bar{Z}^i&=\bar{z}^i+2\pi \bar{p}^i t+{i \over \sqrt{2}}\sum_{n\neq
0}({1\over n}\bar{\beta}_n^i e^{-2\pi i
n(t-\sigma)}+\wt{\bar{\beta}}_n^i e^{-2\pi i n(t+\sigma)})\cr
\lambda^i&=\sqrt{2\pi}\sum_r\lambda_r^ie^{-2\pi i r(t-\sigma)}\cr
\bar{\lambda}^i&=\sqrt{2\pi}\sum_r\wt\lambda_r^ie^{-2\pi i r(t+\sigma)}\cr
\wt\lambda^i&=\sqrt{2\pi}\sum_r\wt\lambda_r^ie^{-2\pi i r(t+\sigma)}\cr
\wt{\bar{\lambda}}^i&=\sqrt{2\pi}\sum_r\wt{\bar\lambda}_r^ie^{-2\pi i
r(t+\sigma)}},}
with commutation relations
\eqn\coommb{\eqalign{
[\beta_n^i,\bar{\beta}_m^j]&=[\wt\beta_n^i,\wt{\bar\alpha}_m^j]
=n\delta_{n+m}\delta_{ij}\cr
\{\lambda^i_r,\bar{\lambda}^j_s\}&=\{\wt\lambda^i_r,
\wt{\bar{\lambda}}^j_s\}=\delta_{r+s}\delta_{ij}}}
with the rest of (anti)commutators vanishing.

When closed strings are in an orbifold background, their
oscillator modding change, since the string is identified when
going once around $\sigma$ via the orbifold group. For the ${\bf
C}^3/{\bf Z}_N$ orbifold with action
\eqn\actbee{\eqalign{ Z^i&\rightarrow e^{2\pi i\nu_i}Z^i\qquad
\b{Z}^i\rightarrow e^{-2\pi i \nu_i}\b{Z}^i\
\cr \lambda^i&\rightarrow e^{2\pi i\nu_i}\lambda^i \qquad
\b{\lambda}^i\rightarrow e^{-2\pi i \nu_i}\b{\lambda}^i},}
with $\nu_1+\nu_2+\nu_3=0(1)$,
the $m$-twisted sector the worldsheet fields have the following
expansion
\vfill\eject
\eqn\closedexp{\eqalign{
Z^i&={i \over
\sqrt{2}}\sum_{n\neq 0}\left({1\over n+m\nu_i}\beta_{n+m\nu_i}^i 
e^{-2\pi i
(n+m\nu_i)(t-\sigma)}+{1\over n-m\nu_i}\wt\beta_{n-m\nu_i}^i 
e^{-2\pi i (n-m\nu_i)(t+\sigma)}
\right)\cr
\bar{Z}^i&={i \over \sqrt{2}}\sum_{n\neq
0}\left({1\over n-m\nu_i}\bar{\beta}_{n-m\nu_i} e^{-2\pi i
(n-m\nu_i)(t-\sigma)}+{1\over n+m\nu_i}\wt{\bar{\beta}}_{n+m\nu_i} 
e^{-2\pi i (n+m\nu_i)(t+\sigma)}
\right)
\cr
\lambda^i&=\sqrt{2\pi}\sum_r\lambda_{r-m\nu_i}^i
e^{-2\pi i (r-m\nu_i)(t-\sigma)}\cr
\bar{\lambda}^i&=\sqrt{2\pi}\sum_{r}\wt\lambda_{r-m\nu_i}^i
e^{-2\pi i (r-m\nu_i)(t+\sigma)}\cr
\wt\lambda^i&=\sqrt{2\pi}\sum_{r}\wt\lambda_{r-m\nu_i}^i
e^{-2\pi i (r-m\nu_i)(t+\sigma)}\cr
\wt{\bar{\lambda}}^i&=\sqrt{2\pi}\sum_r\wt{\bar\lambda}_{r+m\nu_i}^i
e^{-2\pi i
(r+m\nu_i)(t+\sigma)}},}
with the following  commutation relations
\eqn\commutacokp{\eqalign{
[\beta_{n+m\nu_i}^i,\bar{\beta}_{m-m\nu_i}^j]&=
[\wt\beta_{n-m\nu_i}^i,\wt{\bar\alpha}_{m+m\nu_i}^j]
=(n+m)\nu_i\delta_{n+m}\delta_{ij}\cr
\{\lambda^i_{r+m\nu_i},\bar{\lambda}^j_{s-m\nu_i}\}&=
\{\wt\lambda^i_{r-m\nu_i},
\wt{\bar{\lambda}}^j_{s+m\nu_i}\}=\delta_{r+s}
\delta_{ij}}.}

The cylinder amplitude computation in the boundary state formalism can
be performed by using the explicit expressions for the boundary
states we have found and the closed string Hamiltonian in the
$m$-th twisted sector
\eqn\hamiltwist{\eqalign{
H_{c}&=\pi
p^2+2\pi\sum_{\mu=0,1}\left(\sum_{n=1}^{\infty}\am^\mu\ap^\mu
+\sum_{r>0}r\pmi^{\mu}\pp^{\mu}\right)\cr
&+2\pi\sum_{i=1,2,3}\Biggl(\sum_{n=-\infty}^{\infty}\beta^i_{n+m\nu_i}
\bar{\beta}^i_{-n-m\nu_i}+\wt\beta^i_{n-m\nu_i}\wt\beta^i_{n-m\nu_i}\wt{\bar{\beta
}}^i_{-n+m\nu_i}\cr &+\sum_r
(r-m\nu_i)\lambda^i_{-r+m\nu_i}\bar{\lambda}^i_{r+m\nu_i}
+(r+m\nu_i)\wt\lambda^i_{-r-m\nu_i}\wt{\bar{\lambda}}^i_{r+m\nu_i}\Biggl)+
2\pi C_0},}
where $C_0$ is the zero point energy which can be easily computed in the different 
sectors
using the fact that for a complex boson transforming as $e^{2\pi i
a}$ it is $-{1\over 12}+{1\over 2}a(1-a)$ and opposite for a
complex
fermion.
Similar results can be obtained for the ${\bf C}^3/{\bf
Z}_N\times {\bf Z}_N$ orbifold.

In computing open string partition functions and the matrix
elements of the boundary states, it is convenient to introduce the
following functions
\eqn\functions{\eqalign{
\eta(\tau)&=q^{1/24}\prod_{n=1}^\infty(1-q^n)\cr
\vt_1(\nu,\tau)&=2\exp(\pi i \tau/4)\sin(\pi
\nu)\prod_{n=1}^{\infty}(1-q^n)(1-e^{2\pi i \nu}q^n)(1-e^{-2\pi i
\nu}q^n)\cr
\vt_2(\nu,\tau)&=2\exp(\pi i \tau/4)\cos(\pi
\nu)\prod_{n=1}^{\infty}(1-q^n)(1+e^{2\pi i \nu}q^n)(1+e^{-2\pi i
\nu}q^n)\cr
\vt_3(\nu,\tau)&=\prod_{n=1}^{\infty}(1-q^n)
(1+e^{2\pi i \nu}q^{n-1/2})(1+e^{-2\pi i
\nu}q^{n-1/2})\cr
\vt_4(\nu,\tau)&=\prod_{n=1}^{\infty}(1-q^n)
(1-e^{2\pi i \nu}q^{n-1/2})(1-e^{-2\pi i
\nu}q^{n-1/2})}}
In order to compare the answer in the open string channel with the
closed string one, we need the modular properties of $\vt$
functions
\eqn\tranform{\eqalign{
\eta(\tau)&=(-i\tau)^{-1/2}\eta(-1/\tau)\cr
\vt_1(\tau)&=i(-i\tau)^{-1/2}e^{-\pi i \nu^2}\vt_1(\nu/\tau,-1/\tau)\cr
\vt_2(\tau)&=(-i\tau)^{-1/2}e^{-\pi i \nu^2}\vt_4(\nu/\tau,-1/\tau)\cr
\vt_3(\tau)&=(-i\tau)^{-1/2}e^{-\pi i \nu^2}\vt_3(\nu/\tau,-1/\tau)\cr
\vt_4(\tau)&=(-i\tau)^{-1/2}e^{-\pi i
\nu^2}\vt_2(\nu/\tau,-1/\tau)}.}

\listrefs
\end